\documentclass{sigmod}

\usepackage{pslatex,amsmath,epsfig}

\usepackage{amssymb}
\usepackage{cite}
\usepackage{xspace}

\usepackage{multirow}

\usepackage[linesnumbered,ruled,vlined]{algorithm2e}

\usepackage{enumerate}
\usepackage{subfigure}
\usepackage{graphicx}

\newcommand{\eop}{\hspace*{\fill}\mbox{$\Box$}}
\newcommand{\eat}[1]{}
\newcommand{\stitle}[1]{\vspace{0.5ex}\noindent{\bf #1}}

\newcommand{\kw}[1]{{\ensuremath {\mathsf{#1}}}\xspace}
\newcommand{\MatchDist}{\mbox{\sf Match}\xspace}
\newcommand{\AnchorMatchDist}{\mbox{\sf AnchorMatch}\xspace}

\newcommand{\IsMatch}{\mbox{\sf IsMatch}\xspace}
\newcommand{\matchDist}{\mbox{\sf matchDist}\xspace}
\newcommand{\minMatchDist}{\mbox{\sf minMatchDist}\xspace}
\newcommand{\Dist}{\mbox{\sf Dist}\xspace}
\newcommand{\minDist}{\mbox{\sf minDist}\xspace}

\newcommand{\cellbtree}{\mbox{{B$^{\sf ck}$}-tree}\xspace}

\newcounter{example}%[section]
\renewcommand{\theexample}{\arabic{example}}
\newenvironment{example}{
    \refstepcounter{example}
    {\vspace{1ex} \noindent\bf  Example  \theexample:}}{
    \eop\vspace{1ex}} %\hspace*{\fill}\vspace*{1ex}}

\newcounter{defi}
\newenvironment{definition}{
    \refstepcounter{defi}
    {\vspace{1ex} \noindent\bf  Definition  \arabic{defi}:}}{
    \eop\vspace{1ex}} %\hspace*{\fill}\vspace*{1ex}}

\newcounter{prop}
\newenvironment{proposition}{
    \refstepcounter{prop}
    {\vspace{1ex} \noindent\bf  Proposition  \arabic{prop}:}}{
    \eop\vspace{1ex}} %\hspace*{\fill}\vspace*{1ex}}

\newcounter{lemma}%[section]
\newcommand{\thetheorem}{\arabic{lemma}}
\newenvironment{lemma}{\begin{em}
    \refstepcounter{lemma}
    {\vspace{1ex} \noindent\bf  Lemma  \thetheorem:}}{
    \end{em}\eop\vspace{1ex}} %\hspace*{\fill}\vspace*{1ex}}
\renewenvironment{proof}{
    {\vspace{1ex} \noindent\bf Proof Sketch:}}{
    \eop \vspace{1ex} }

\begin{document}

\title{Efficient Spatial Keyword Search in  Trajectory Databases}
%
% You need the command \numberofauthors to handle the "boxing"
% and alignment of the authors under the title, and to add
% a section for authors number 4 through n.
%

%%%%%%%%%%%

%\numberofauthors{2}

\author{%
{ Gao Cong$~^\dag$  \hspace*{20pt} Hua Lu$~^\S$  \hspace*{20pt} Beng Chin Ooi$~^\ddag$ \hspace*{20pt} Dongxiang Zhang$~^\ddag$\hspace*{20pt} Meihui Zhang$~^\ddag$  }%
\vspace{1.6mm}\\
\fontsize{10}{10}\selectfont%\itshape
$~^\dag$School of Computer Engineering, Nanyang Technological University \\
\fontsize{10}{10}\selectfont$~^\S$Department of Computer Science, Aalborg University\\
\fontsize{10}{10}\selectfont$~^\ddag$School of Computing, National University of Singapore \\
\fontsize{9}{9}\selectfont\ttfamily\upshape
gaocong@ntu.edu.sg\hspace*{15pt} luhua@cs.aau.dk\hspace*{15pt}  \{ ooibc |  mhzhang | dxzhang\}@comp.nus.edu.sg%
}
%%%%%%%%%%%

\maketitle

\pagenumbering{arabic}

\begin{abstract}

An increasing amount of trajectory data is being annotated with
text descriptions to better capture the semantics associated
with locations. The fusion of spatial locations and text
descriptions in trajectories engenders a new type of top-$k$
queries that take into account both aspects. Each trajectory in
consideration consists of a sequence of geo-spatial locations
associated with text descriptions. Given a user location
$\lambda$ and a keyword set $\psi$, a top-$k$ query returns $k$
trajectories whose text descriptions cover the keywords $\psi$
and that have the shortest match distance. To the best of our
knowledge, previous research on querying trajectory databases
has focused on trajectory data without any text description,
and no existing work has studied such kind of top-$k$ queries
on trajectories.
This paper proposes one novel method for efficiently computing
top-$k$ trajectories. The method is developed based on a new
hybrid index,
cell-keyword conscious B$^+$-tree, denoted by \cellbtree,
which enables us to exploit both text relevance and location
proximity to facilitate efficient and effective query processing.
The results of our extensive empirical studies with an
implementation of the proposed algorithms on BerkeleyDB
demonstrate that our proposed methods are capable of achieving
excellent performance and good scalability.

\end{abstract}

%==========================================
\section{Introduction}
%==========================================

With the increasing popularity of crowdsourcing, as well as the
advancements and miniaturization of handheld devices with GPS
receivers, massive amount of data that are geo-tagged or
associated with text information are being generated at an
unprecedented scale.
For example, crowdsourcing of motion trajectories is applied to
generate the Open map systems (e.g., openstreetmap.org and
waze.com).

\begin{figure}[!t]
 \begin{center}
  \includegraphics[width=0.8\columnwidth]{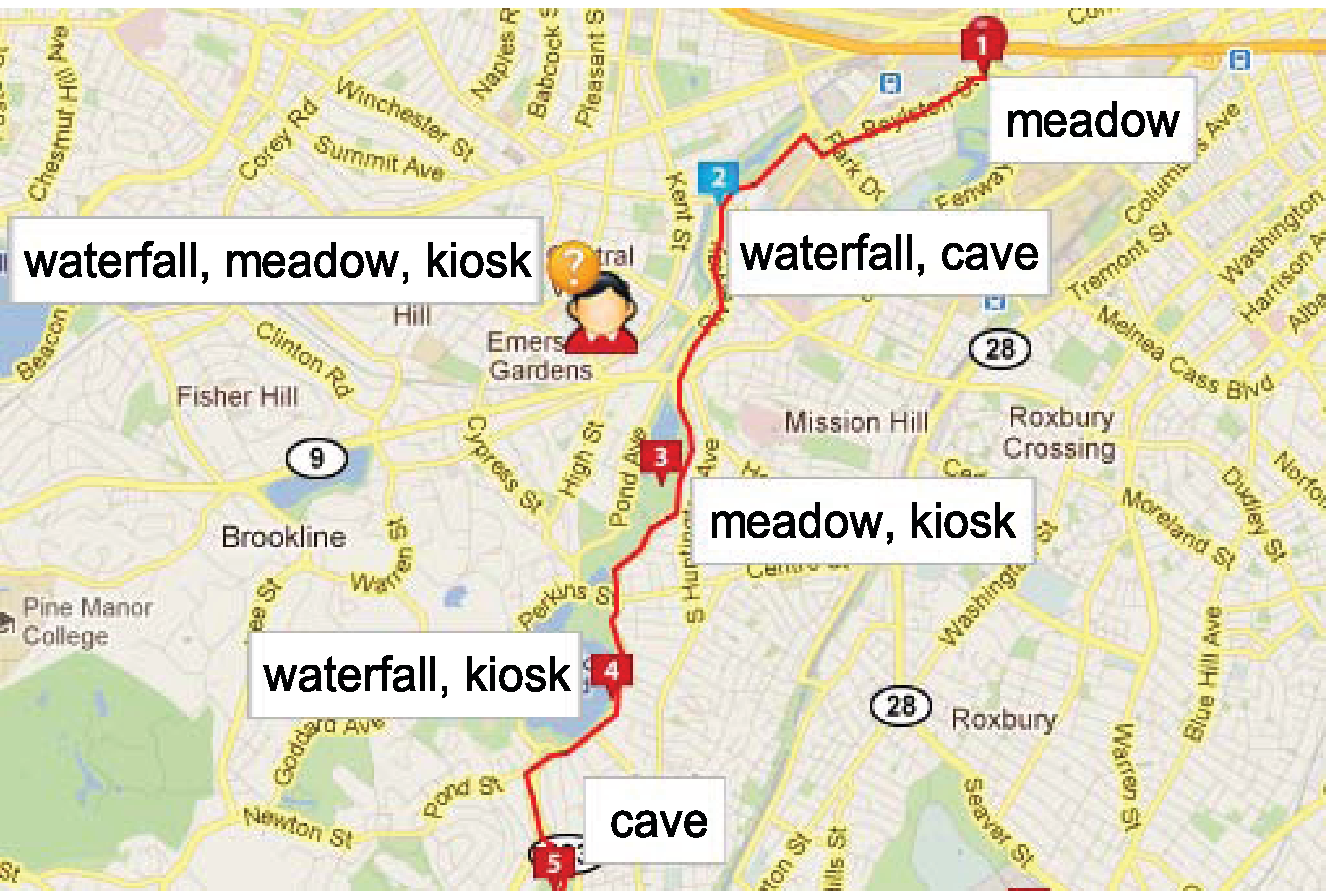}
    %\vspace{-2ex}
  \caption{Example}\label{fig:example}
  \end{center}
  \vspace{-2ex}
\end{figure}

Users have crowdsourced huge volumes of trajectory data that are
annotated with keywords or text descriptions. In such datasets, a
trajectory is composed of a sequence of places and line segments
connecting these places.
The places in a trajectory, captured as spatial locations, are
often associated with text descriptions.
Figure~\ref{fig:example} shows an example of a trajectory.
%The following describes for example sources of such trajectories.
Such trajectories come from various sources, and we name just a
few in the following:
1)~In many GPS-trajectory-sharing websites (e.g., Mountain
Bike: www.bikely.com,   GPS sharing: www.gpssharing.com,
GPSies: www.gpsies.com, and
Geolife~\cite{DBLP:journals/debu/ZhengXM10}), people upload
their travel routes. To record their journeys or share life
experiences with others, they often attach texts and multimedia
content (e.g., photos) as annotations to the places in their
trajectories.
2)~In location-based social network services (e.g.,
FourSquare), each place is associated with tags and users can
check in such places.
The check-in sequence of a user in a period forms a trajectory.
The places can points of interests of any kind, e.g.,
restaurants, shops, and thus, the trajectories can be of
various types, such as travel trajectories and daily life
trajectories.
3)~Trajectories with text descriptions can be extracted from
travel itineraries~\cite{HaoCWXYPZ10},
% in which each trajectory
%contains a sequence of places, and each of them is associated with
%text description.
%
as well as Flickr photos~\cite{Lu:2010}.
% in which each trajectory contains a sequence
%of places, and each of them is associated with text description (and
%tags and photos).
%

Such publicly accessible datasets serve as an informative repository
to users. A user may want to find others' travel routes that are
relevant to his/her interests and that have a short travel distance.
Motivated as such, we consider queries that search previously
explored routes of places that satisfy a user's interests or needs,
expressed as a set of keywords, and that may also lead to the
shortest total traveling distance.
The results of such a query exploit the collective intelligence of
crowdsourcing.
%
% move to the related work
%Note that such a query is complementary to the route planning
%queries~\cite{Sharifzadeh:2008}, which return a route of places from
%a spatial database such that the route covers a set of query
%keywords and the travel distance is minimized.

In addition, users may be interested in learning the daily life
experience of others. For example,
 from relevant social network applications, it is easy to derive a shopping trajectory
database, where each place (corresponding to a shop) in a
user-generated trajectory is associated with the items bought
by the user at that place. Such a user-generated trajectory
indicates the user's preferences.
Suppose that a user has a shopping list of product names. She
would like to see the routes of other users who buy all the
items on the list, and the traveling distance from her starting
location along this route that is the minimum.

The fusion of spatial locations and text descriptions in
trajectories demands efficient processing of queries that
involve both attributes.
Indeed, the aforementioned GPS sharing websites already support a
type of queries related to both text and locations, namely the
keyword range queries, to help users share, browse and search GPS
trajectories. They allow users to specify a region and a set of
keywords, and return the trajectories that are inside the query
region and contain the set of query keywords.
However, the algorithms used are not publicized, and the response
for answering such queries in these websites is very slow.
%
%2)~search by region tags. For example, a users can search using
%keywords containing locations, e.g., finding trajectories containing
%keywords NYC. This however will return a huge number of
%trajectories.

Existing research on querying trajectory database has focused
on trajectory data without any text description. For example, a
$k$-Nearest Neighbor query~\cite{DBLP:conf/ssd/FrentzosGPT05}
returns the $k$ nearest moving object trajectories to a given
query point based on the minimum distance from the query point
to a trajectory.
Querying trajectory data is time consuming and therefore, indexes
such as the R-tree and its optimized versions for trajectories
%,e.g., the TB-tree~\cite{PfoserJT00},
have been used.

To the best of our knowledge, no publication considers querying
trajectories that are composed of a sequence of geo-locations
associated with text descriptions.

In this paper, we introduce a new problem: the top-$k$ spatial
keyword query (T$k$SK) on trajectories. Given a large database of
trajectories, a T$k$SK query consists of a spatial element (query
location) and a set of keywords, and it returns the top-$k$
trajectories with the shortest match distance. The match distance is
measured by the sum of two distances: the length of a sub-trajectory
covering all query keywords, and the distance from the query
location to the start location of the sub-trajectory.
It is a challenge to efficiently answer the T$k$SK query on
trajectories associated with text.

To this end, we propose a novel solution with the following
features.
First, we develop a new index for trajectories, called cell-keyword
conscious B$^+$-tree, denoted by \cellbtree. \cellbtree integrates
spatial information captured by location keys generated by adaptive
cells and text information such that it enables simultaneous
application of both spatial proximity and keyword matching in query
processing.
%
%Unlike existing approaches that use R-tree based indexes for
%indexing spatial-keyword Web objects~\cite{vldb09}, we use the
%B$^+$-tree to index trajectories associated with keywords.
%
The \cellbtree is efficient for queries as well as updates, and
it is adaptive to varying workloads. Further, with the use of
the B$^+$-tree that is available in all mainstream DBMSs, our
proposed solution can be easily grafted onto existing database
systems.
Second, based on the \cellbtree, we develop an algorithm for
choosing candidate trajectories that are close to the query location
and contain the query keywords, and thus are more likely to be the
results of a T$k$SK query.
Third, we propose a linear time algorithm, called Match, for
efficiently computing the match distance between a query and a
candidate trajectory, which contrasts with a straightforward
method that takes quadratic time.

%A salient feature of the proposed solution is that it is ready
%to be implemented in DBMSs, and we employ BerkeleyDB in our
%work.

Since no baseline algorithms exist for processing T$k$SK
queries, we also develop four baseline algorithms. They all use
the proposed algorithm Match for computing the matching
distance. They differ in their ways of finding candidate
trajectories:~1)~The first one uses the inverted list index to
choose the trajectories containing all query words.
%, and then employs algorithm Match for computing the matching
%distance.
2)~The second uses the R-tree to retrieve nearby trajectories.
%and uses the proposed algorithm for computing the match distance.
3)~The third is based on the IR-tree~\cite{vldb09}, treating
each trajectory as a whole to retrieve nearby trajectories
containing query keywords.
4)~The fourth extends the TB-tree~\cite{PfoserJT00}, an
existing index for trajectories, to incorporate the text
information organized in an inverted index, and uses the
extended TB-tree to retrieve candidate trajectories.

%Note that the second baseline indexes the places of trajectories
%while the first baseline indexes the whole trajectories. Both
%strategies are different from the proposed method that actually
%indexes trajectory segments together with associated texts.

In summary, the paper's contributions are threefold. First, we
introduce and formalize a new type of queries on trajectory data
that are associated with words. Second, we propose a novel solution
for efficiently processing T$k$SK queries. The proposed solution
consists of a new index structure \cellbtree for trajectories
associated with words, an approach to computing the minimum match
distance between a trajectory and a query, and a top-$k$ query
processing algorithm. The proposed solution can be implemented on
top of existing DBMSs cost-effectively.
%The proposed solution is ready to be implemented on top of existing DBMSs.
%
We also explore other ways of answering T$k$SK queries as baseline
methods.
Third, with an implementation of the \cellbtree based algorithm
on BerkeleyDB, we conduct an extensive experimental study,
which includes a comparison with the four baselines. The
experimental results demonstrate the efficiency and scalability
of our proposed solution.

The rest of the paper is organized as follows.
Section~\ref{sec:pro_def} defines the T$k$SK query.
Section~\ref{sec:baseline} presents the baseline algorithms.
Section~\ref{sec:alg} details our solution for processing the T$k$SK
query. Section~\ref{sec:exp} reports the experimental study.
Section~\ref{sec:related} reviews related work.
Section~\ref{sec:conc} concludes this paper. % and offers future directions.
%Baseline algorithms are presented in the Appendix.

%==========================================
\section{Problem Statement}\label{sec:pro_def}
%==========================================
In this section, we give the problem statement and provide
necessary definitions and background.

\stitle{Data}
%
% a sequence of locations in which each location has text + position
%
Let $\mathcal{D}$ be a dataset in which each object is a trajectory.

\begin{definition} Trajectory\\
Each trajectory ${\mathcal {TR}} \in \mathcal{D}$ is defined as a
sequence of places (points of interests) ${\mathcal {TR}}
  = \kw{L}_{1}, \cdots , \kw{L}_{i}, \cdots , \kw{L}_{n}$.
\end{definition}

Each place $\kw{L}$ is represented by a pair $(\kw{L}.\lambda,
\kw{L}.\psi)$, where $\kw{L}.\lambda$ represents a geo-spatial point
location and $\kw{L}.\psi$ denotes a set of keywords (e.g., the
description about the place).

We denote the union of the text description of each place in
trajectory $\mathcal {TR}$ by ${\mathcal {TR}}.\psi$ = $\bigcup_{i
=1}^{n} \mathcal {TR}.\kw{L}_i.\psi$.

\begin{definition} Sub-Trajectory and Contain\\
We define a sub-trajectory as a subsequence from place $s$ to place
$e$ of trajectory $\mathcal {TR}$ as $\mathcal {TR}.\kw{L}_{s}^{e}$,
$s, e \in [1, n], s \leq e $. Given two sub-trajectories $\mathcal
{TR.L}_{s1}^{e1}$ and $\mathcal {TR}.\kw{L}_{s2}^{e2}$, we say that
$\mathcal {TR}.\kw{L}_{s1}^{e1}$ \emph{contains} $\mathcal
{TR.L}_{s2}^{e2}$ if $s1 \leq s2$ and $e1 \geq e2$.
\end{definition}

We denote the union of the text description of each place in
sub-trajectory $\mathcal {TR}_{s}^{e}$ by ${\mathcal
{TR}}_{s}^{e}.\psi$ = $\bigcup_{i =s}^{e} \mathcal
{TR}.\kw{L}_i.\psi$.
%It is also allowed that ${\mathcal {TR}}$ is associated with text
%about the whole trajectory.

\stitle{Query}
A spatial keyword query  $q=\langle \lambda, \psi \rangle$ has two
components, where $q.\lambda$ is a spatial location and $q.\psi$ is
a set of keywords. The location descriptor $q.\lambda$ specifies the
location preference of a user, and $q.\psi$ indicates the preference
of a user on the keywords of objects.

\begin{definition} Match\\
We say that a trajectory $\mathcal{TR}$ \textbf{matches} a query $q$
if the following condition is satisfied.
$$q.\psi \subseteq \mathcal{TR}.\psi$$
Similarly, we say that a sub-trajectory $\mathcal
{TR}.\kw{L}_{s}^{e}$ matches a query if $q.\psi$ $\subseteq$
$\mathcal {TR}_{s}^{e}.\psi$. \label{def:match}
\end{definition}

Intuitively,  we say that a trajectory $\mathcal{TR}$
\textbf{matches} a query $q$ if all the keywords of the query are
contained in the text of the trajectory.

\begin{definition} Minimum Match\\
We say that a sub-trajectory $\mathcal {TR}.\kw{L}_{s}^{e}$ is a
minimum match of a query $q$ if (1) $\mathcal {TR}.\kw{L}_{s}^{e}$
matches $q$; and (2) no sub-trajectory of $\mathcal
{TR}.\kw{L}_{s}^{e}$ matches $q$.
\end{definition}

\begin{example}
Refer to Figure~\ref{fig:example}. A traveller wants to find a route
in which she can see waterfall, panda and kiosk. There are two
minimum matches to the query \{\emph{waterfall}, \emph{meadow},
\emph{kiosk}\}: $\kw{L}_2$$->$$\kw{L}_3$ and
$\kw{L}_3$$->$$\kw{L}_4$.
Note that  $\kw{L}_2$$->$$\kw{L}_3$$->$$\kw{L}_4$ is a match but it
is not a minimum match.
\end{example}

\begin{definition}\label{def:matchDist} Match Distance\\  %$\mathrm{matchDist}$($q, \mathcal{TR}$)\\
If a sub-trajectory $\mathcal{TR.L}_{s}^{e}$ \textbf{matches} a
query $q$, the match distance $\mathrm{matchDist}$($q,
\mathcal{TR.L}_{s}^{e}$) is defined as follows:

\begin{displaymath}
\begin{aligned}
&{\matchDist}(q, \mathcal{TR}.\kw{L}_{s}^{e}) = \min\{ \Dist(q,
\mathcal{TR}.\kw{L}_s),
\Dist(q, \mathcal{TR}.\kw{L}_e)\}\\
& + \sum_{i=s}^{e-1} \Dist(\mathcal{TR}.\kw{L}_i, \mathcal{TR}.\kw{L}_{i+1})),
%&s.t., \mathcal {TR}.\kw{L}_{s}^{e} \mbox{is a  match of } q.
\end{aligned}
\end{displaymath}

where $\Dist$(.,.) is the Euclidean distance between two
locations.

If a sub-trajectory $\mathcal{TR.L}_{s}^{e}$ does not {\it match} a
query $q$, the match distance is defined as $\infty$.
%if they are two geographical points, or the minimum Euclidean
%distance between them if they are two geographical regions.
\end{definition}

\begin{definition} Minimum Match Distance\\  %$\mathrm{matchDist}$($q, \mathcal{TR}$)\\
If a trajectory $\mathcal{TR}$ \textbf{matches} a query $q$, the
minimum match distance $\minMatchDist$($q, \mathcal{TR}$) is defined
as follows:
\begin{displaymath}
\begin{aligned}
&\minMatchDist(q, \mathcal{TR})= \min_{(s,e)} (\matchDist(q, \mathcal{TR}.\kw{L}_{s}^{e})),&\\
%&\min_{(s,e)} (\matchDist(q, \mathcal{TR}.\kw{L}_{s}^{e})),\\
&s.t., \mathcal{TR}.\kw{L}_{s}^{e} \mbox{is a minimum match of } q.
\end{aligned}
\end{displaymath}
\end{definition}
%
%where $\mathrm{dist}$(.,.) is the Euclidean distance between two
%locations.
% if they are two geographical points, or the minimum
%Euclidean distance between them if they are two geographical
%regions.
%,${\mathcal {TR}.L}_{s}^{e}$ matches $q$.

\begin{example}
Consider the example in Figure~\ref{fig:example}. When we take her
current location $q.\lambda$ into account, sub-trajectory
$\kw{L}_3$$->$$\kw{L}_4$ is the one with the minimum Match Distance.
\end{example}

\begin{definition} {\em Top-$k$ Spatial Keyword query }(T$k$SK)\\
%The result of the {\em top-$k$ spatial keyword query} $q$, denoted
%by $\mathit{Ans(q)}$, is a subset  of $\mathcal{D}(q.\psi)$ with $k$
%trajectories such that $\forall t \in \mathit{Ans(q)} $\;
%%
%( $ \forall t' \in \mathcal{D}-\mathit{Ans(q)}$ $(
%\mathrm{matchDist}(q.\lambda,t.\lambda) \le
%\mathrm{matchDist}(q.\lambda,t'.\lambda))$).
%
Given a trajectory set $\mathcal{D}$, a {\em top-$k$ spatial keyword
query} (T$k$SK) with $q=\langle \lambda, \psi\rangle$ returns from
$\mathcal{D}$ $k$ trajectories that have the smallest minimum match
distances with respect to $q$, each associated with the start and
end place indexes that yield the minimum match  distance.
%
%Suppose that each trajectory
%$\mathcal{TR}_i$ is identified by its id $i$.

%%%%%%% ooibc2:

Formally, a T$k$SK query returns a set $\mathit{Ans(\mathcal{D},
q)}$ of $k$ triples ($t, s, e$), where $t \in \mathcal{D}, 1\leq s
\leq e \leq |t|$, such that
\begin{enumerate}
  \item $|\mathit{Ans(\mathcal{D}, q)}| = |\pi_1(\mathit{Ans(\mathcal{D}, q)})|=k$, where  $\pi_1(.)$ denote the projection on the first attribute of a set of triples of the format $(t, s, e)$.
  \item $\forall (t, s, e) \in Ans(\mathcal{D}, q)$,
$\matchDist(q,t.\kw{L}_s^e) = \minMatchDist(q,t)$;
  \item $\forall (t, s, e) \in Ans(\mathcal{D}, q)$,
  $ \forall t' \in \mathcal{D}
\setminus \pi_1(\mathit{Ans(\mathcal{D}, q)})$, the following inequality holds: $\minMatchDist(q,t) \le \minMatchDist(q,t')$.
\end{enumerate}

Intuitively, the answer to the query consists of $k$
sub-trajectories from $k$ distinct trajectories whose minimum match
distances to query $q$ are the smallest.
%$\forall (i,s,e)\in \mathit{Ans(\mathcal{D}, q)}\; (\not\exists
%(i',s',e')\in \mathbb{N} \times \mathbb{N} \times \mathbb{N}\; ({\rm
%matchDist}(q, \mathcal{TR}_i.\kw{L}^s_e) > {\rm matchDist}(q,
%\mathcal{TR}_{i'}.\kw{L}^{s'}_{e'})))$.
%
\end{definition}

%The answer to query $q$ is a list of the $k$ sub-trajectories with
%the smallest match distance to query $q$.

%%%%%%%%%%%%%%%%%%%%%%%%%%%%%%%%%%%%%%%%%%%%%%%%%%%%%%%%%%%%%%%%
\section{Baseline Algorithms} \label{sec:baseline}
%%%%%%%%%%%%%%%%%%%%%%%%%%%%%%%%%%%%%%%%%%%%%%%%%%%%%%%%%%%%%%%%

No baseline algorithm exists for the top-$k$ spatial keyword queries
on trajectory data. We develop four baseline algorithms. The four
baseline algorithms constitute a contribution to the problem of
processing the top-$k$ spatial keyword queries in that they explore
the possibility of using existing index techniques for the new
problem.
The four baseline algorithms employ the algorithm (presented in
Section~\ref{sec:alg:match}) for computing match distance. Baseline
4 is lengthy and is described in Appendix.
The baseline algorithms act as background for better
understanding of the problem and its complexity.

\subsection{Baseline 1: IF}
The first baseline, IF, uses Inverted File as the index
structure. Specifically, it aggregates the text description
associated with each place in a trajectory to get a set of
words of the trajectory, and then builds inverted file for all
the trajectories.

The idea of the IF algorithm is to use the inverted file to
filter out the trajectories that do not contain all the
keywords of query $q$, {\it i.e.}, finding the set of
trajectories $T_m$ that match the query. Then for each
trajectory in $T_m$, we compute its $\mathrm{matchDistance}$ to
the query using the algorithm presented in
Section~\ref{sec:alg:match} and find the top-$k$ trajectories.

\subsection{ Baseline 2: RT}
The second baseline, RT, uses an R-tree~\cite{Antonin84} as the
index structure.
%and is called RT. Note that the R-tree and its
%variants are the dominating structure for indexing
%trajectories.
Specifically, it aggregates the MBR associated
with each place in a trajectory to get the MBR of the
trajectory, and then uses an R-tree to index all the
trajectories. For each trajectory, this baseline uses a
separate index structure to organize the text description
associated with places of the trajectory as the component 2 in
Section~\ref{sec:alg:index}.

Given a query $q$, the baseline uses the R-tree to find the nearest
trajectory incrementally. For each nearest trajectory, we check if
it matches the query keywords. If yes, we compute its
$\mathrm{matchDistance}$ to the query using the algorithm in Section
~\ref{sec:alg:match}.
In the process, the algorithm keeps track of the minimum match
distance of the current $k$th trajectory, denoted by
$\mathit{threshold}$.
For a newly ``seen'' trajectory with spatial distance
$\mathit{dist}$ to query $q$, if the score $\mathit{dist}$ exceeds
$\mathit{threshold}$, the algorithm stops since it is guaranteed
that all ``unseen'' trajectories will not have smaller match
distance than the current $k$'th trajectory (and thus cannot be in
the result). Note that $\mathit{dist}$ is a lower bound of the
minimum match distance.

\subsection{  Baseline 3: IRT}
The third baseline, IRT, uses the IR-tree~\cite{vldb09} as the
index structure, which is used to index spatial Web objects.
The IR-tree is essentially an R-tree~\cite{Antonin84} extended
with inverted files~\cite{Zobel}
%(an efficient index for text retrieval).
%
Each leaf node in the IR-tree contains a number of entries of the
form $(p, p.\lambda)$, where $p$ refers to the identifier of a
spatial object, and $p.\lambda$ is the bounding rectangle of $p$.
Each leaf node also contains a pointer to an inverted file for the
text descriptions of the objects stored in the node.
Each non-leaf node $R$ in the IR-tree contains a number of entries
of the form $(cp, rect, cp.di)$ where $cp$ is the address of a child
node of $R$, $rect$ is the MBR of all rectangles in entries of the
child node, and $cp.di$ is the identifier of a pseudo text
description of the child node. The pseudo text description is a
union of all text descriptions in the entries of the child node.
Each non-leaf node also contains a pointer to an inverted file for
the pseudo text descriptions of its child nodes.
The pseudo text description enables us to prune a node (and the
subtree under the node) if it does not cover all the query keywords.

To use the IR-tree~\cite{vldb09} to organize the trajectories, we
 aggregate the MBR associated with each place in a
trajectory to get the MBR of the trajectory; similarly we get the
set of words of the trajectory by aggregating the text description
of each place.

%1)~the MBR of a trajectory can be much larger than the real
%geographical space of places in the trajectory, and thus ${\sf
%minDist}(q, \mathcal{TR})$ is not a tight lower bound for match
%distance.

We adapt top-k algorithm presented in~\cite{vldb09} that is based on
the best-first search to find the top-k trajectories. A priority
queue $U$ is used to keep track of the nodes and trajectories that
have yet to be visited. The values of ${\sf minDist}(q, .)$, which
is the minimum Euclidian distance between $q$ and a trajectory (or a
node), are used as the keys.
Note that the key used for a trajectory in $U$ is not the match
distance,
% for ranking trajectories, while the key for an object used
%in the top-k algorithm~\cite{vldb09} is the value of the ranking
%function.
%
but a loose lower bound of the match distance between query and
trajectories in a node. It is used to choose which node to visit
next and when to terminate the algorithm.
When deciding which node to visit next, the algorithm picks the node
$CN$ with the smallest $\minDist(q,CN)$ value in the set of all
nodes that have yet to be visited. The algorithm terminates when the
match distance of $k$th trajectory is smaller than the key of first
element in $U$. Algorithm ~\ref{alg:irt} shows the pseudo-code.

\subsection{ Discussion}

The first baseline IF uses the text information to prune the search
space without utilizing the spatial information to speed up. The
second baseline RT uses the spatial information to guide the search
for results without utilizing the text information.

Different from the first two baselines, the baseline IRT (and the
baseline given in Appendix) is able to make use of both text
information and distance information to prune the search space.
However IRT faces the following challenges: The MBR of a trajectory
can be much larger than the real geographical space of places in the
trajectory, and thus the MBRs of nodes in the IR-tree have large
overlapping.
The text description of a trajectory is the aggregation of the
descriptions of all places in the trajectory.
Hence, the overlapping of text descriptions between nodes with large
overlapping MBRs is also large. Thus the pruning power of the text
information associated in the IR-tree nodes might be limited.

\begin{algorithm}[!t]\small
\caption{IRT ( query $q$, Tree root $root$, Integer
$k$)}\label{alg:irt}
        $V \leftarrow$ new max-priority queue of $k$ elements of $\infty$\; %\Comment{maintain the top $k$ trajectories}
        $U \leftarrow$ new min-priority queue\;
        $U$.Enqueue($root,0$)\;
        \While{$U$ is not empty}
        {
            $e$ $\leftarrow$ $U.$Dequeue()\;
            \If{(\minDist($q.\lambda,e.\lambda$) $\geq$ $V[k]$)} {{\bf break}  while-loop;}
            \If {$e$ is a trajectory}
                {update $V$ by $(e,
                \mathsf{Match}(q, e, V[k]))$\;}
            \Else (\tcp*[f]{$e$ points to a child node})
                {
                read the node $CN$ of $e$\;
                %\State visit the inverted file of $CN$
                read the posting lists of $CN$ for keywords in $q.\psi$\;
                \For{each entry $e'$ in the node $CN$}
                {
                    \If {$q.\psi \subseteq e'.\psi$ and \minDist($q.\lambda,e'.\lambda$) $<$ $V[k]$}
                    {
                        $U$.Enqueue($e',\minDist(q.\lambda,e'.\lambda)$)\;
                    }
                }
            }
        }
    %\EndFor
    {\bf return} $\{ V \}$; \tcp{top-$k$ results }
\end{algorithm}

%\begin{example}
%An example about the algorithm
%\end{example}

%==========================================
\section{Proposed Algorithms for Query Processing} \label{sec:alg}
%==========================================

Section~\ref{sec:alg:index} presents the proposed index \cellbtree.
Based on the index, we present the incremental expansion algorithm
for finding candidate trajectories in
Section~\ref{sec:alg:incremental}.
Section~\ref{sec:alg:match} presents an algorithm for matching a
candidate trajectory with a query.

%==========================================
\subsection{Proposed Index: \cellbtree} \label{sec:alg:index}
%==========================================

\begin{figure}[!t]
 %\begin{minipage}[b]{2in}
 \centering
 \subfigure[Location ID of Cell]{ \hspace{-3ex}
  \includegraphics[width=0.5\columnwidth]{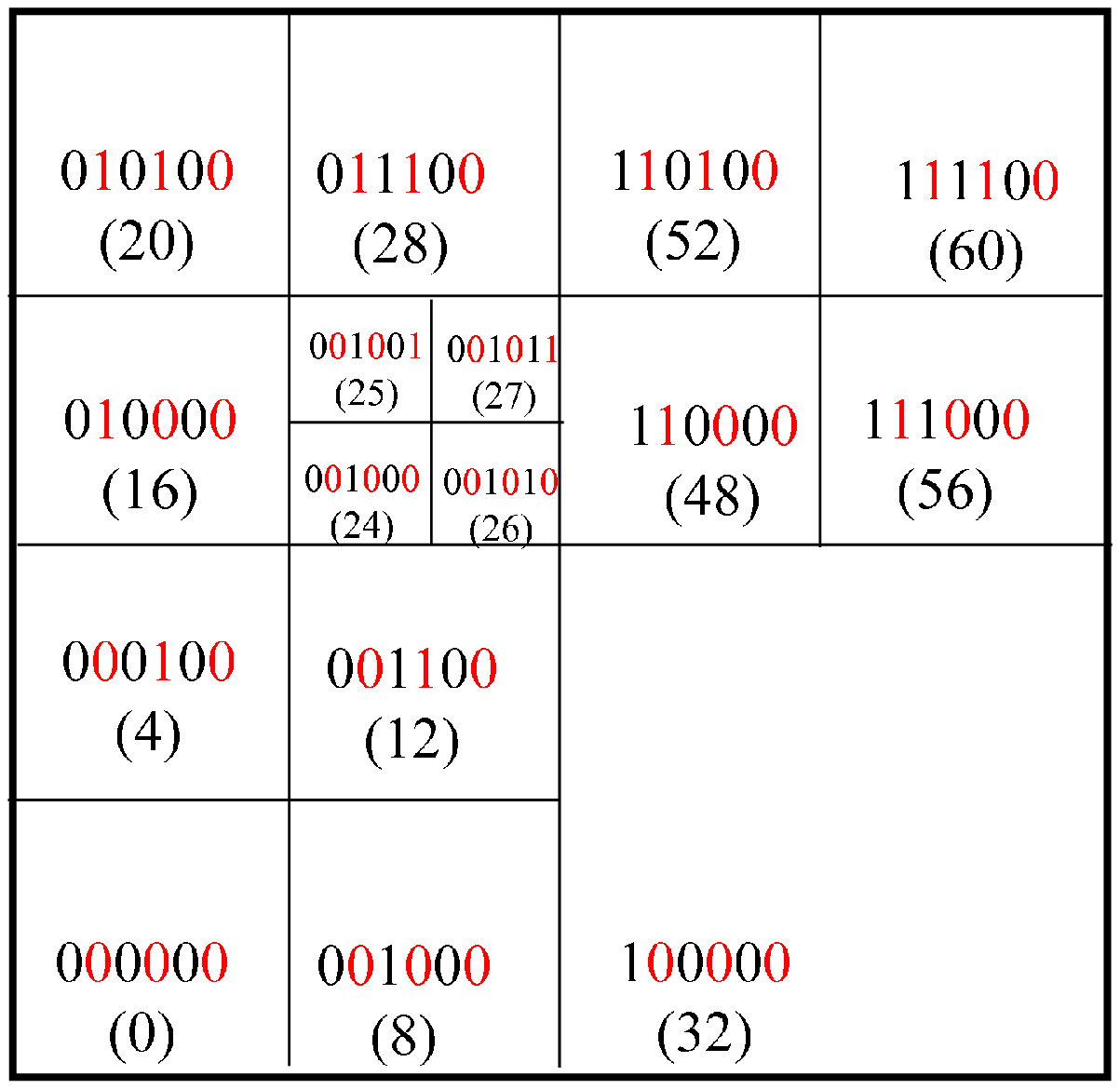}
    %\vspace{-2ex}
  \label{fig:quadcode}
  }
 %\begin{minipage}[b]{2in}
 \subfigure[Three Trajectories]{
  \includegraphics[width=0.5\columnwidth]{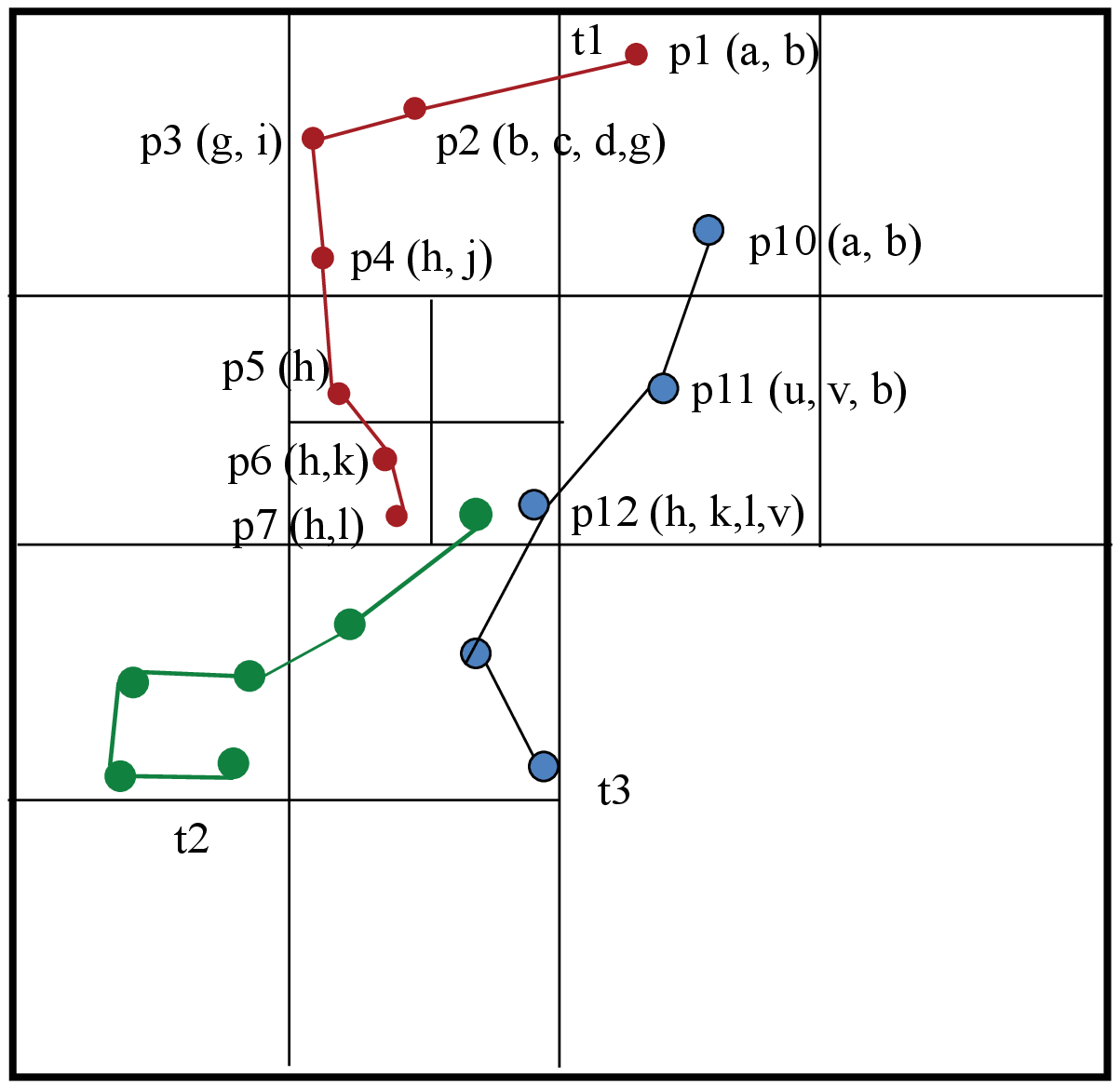}\hspace{-3ex}
    %\vspace{-2ex}
  %\caption{Example for three trajectories}
  \label{fig:quadcode:data}
  %\end{minipage}
%  \vspace{-2ex}
  %\begin{minipage}[b]{2.8in}
  }
  \subfigure[Index]{
 \includegraphics[width=0.8\columnwidth]{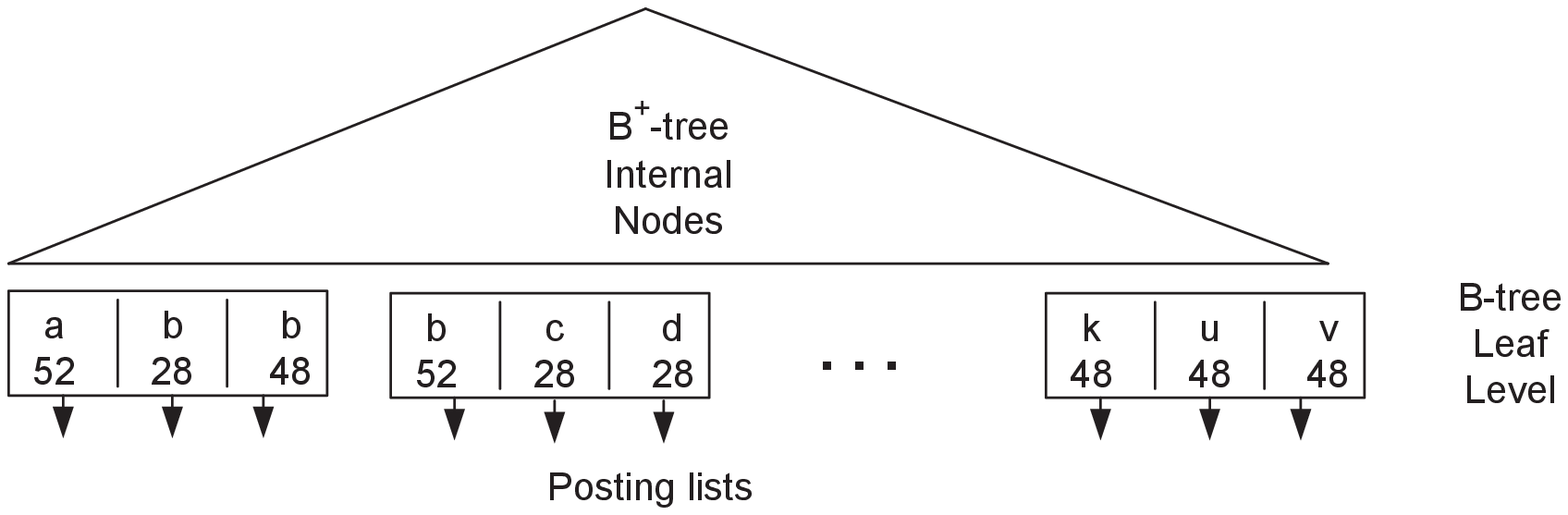}
 \label{fig:quadcode:index}
 }
  \caption{Example }
 %\end{minipage}
  \vspace{-3ex}
\end{figure}

Ideally, we can index trajectories associated with text information
to enable pruning search space by utilizing both spatial distance
and keyword information for efficient query processing.
It is, however, challenging to develop indexes to meet the
complexities of trajectories associated with text information. To
this end, we propose an index, called cell-keyword conscious
B$^+$-tree, denoted by \cellbtree, which comprises two components.

1)~Component 1 is used to locate the IDs of trajectories that are
close to the query location and contain all the keywords. It is used
to organize the segment-level information of trajectories.

2)~Component 2 is used to compute the minimum match distance of a
selected trajectory to query $q$. It is used to organize the
detailed information of each trajectory.

\emph{Component 1}: We divide the spatial region of dataset
$\mathcal{D}$ into quad cells of various sizes to generate location
codes.
The ID of each cell can be generated by using the bit-interleaving
method~\cite{abel83}. %\cite{Ramsak:2000}, which is developed for uniform cells.
If a quad cell consists of a set of uniform cells, the minimum ID of
the set of cells will be the ID of the quad cell.
Figure~\ref{fig:quadcode} shows an example.
Based on the cell division, we build a B$^+$-tree to index
trajectories together with their text descriptions. Each leaf entry
contains three elements:

\begin{itemize}\itemsep=-3pt
\item wordID: it denotes the ID of a word in the trajectory database.

\item cellID: it denotes the ID of a cell that contains a wordID.

\item posting list: it is a sequence
of trajectory identifiers for each wordID  and cellID, {\it i.e.},
the list of trajectories in cell cellID that contain word wordID.
\end{itemize}

In the index, the entries are organized first by the word ID, and
next by the cell ID. Hence, the posting lists for the same word are
organized together, and posting lists of nearby cells for the same
word are together. This enables visiting nearby cells for a word by
following the pointers between leaf nodes of B$^+$-tree.

All distinct words in the text description of the trajectory
database constitute a vocabulary, and each word has a wordID. We
proceed to explain the other two elements, cellID and posting list.

{\bf cellID:} The cellID element aims to integrate the spatial
information and text information of trajectories. We partition the
index space into  cells, and thus one trajectory may span multiple
cells. The sizes of  cells are not fixed. We set the size of a cell
such that the number of trajectories in a cell is smaller than a
threshold $\xi$. Note that the empty cells are not indexed.

%%%%%%%%%%%%BEGIN EAT
\eat{
we take into account the following  factors. 1) The number of
fragments generated by a partitioning. In general we want to avoid
generating too many fragments. 2) The word distribution by a
partitioning

One possible way to recursively partition a space into 4 cells: 1)
We create a $2^m \times 2^m$ grid.
2) for each trajectory, we assign it to grid cells that contain part
of the trajectory or the whole trajectory.
3) from bottom to up, we check each cell to see whether we need to
split it, or whether its three relevant quad cells should be
combined to a higher-level quad cell

}
%%%%%%%%%%%%END EAT

{\bf posting list:}
Given a set of query words, we need to check if a cell contains a
trajectory that covers all the query keywords.
To meet the need, for each wordID $w$, and cellID $c$, a posting
list is a sequence of identifiers of the trajectories such that part
of the trajectory or the whole trajectory falls in cell $c$, and the
text description associated with the trajectory segment in $c$
contains word $w$.
Such a design enables associating the cell ID, which represents the
spatial information of a trajectory segment, with the text
information of the segment.

\begin{example}
In Figures~\ref{fig:quadcode}-\ref{fig:quadcode:data}, for cell 52,
we generate one entry ($a$, 52, $<$t1,t3$>$) since the fragments of
trajectories $t1$ and $t3$ in cell 52 contain word $a$. Similarly,
we generate another entry ($b$, 52, $<$t1, t3$>$) for cell 52. As
another example, for cell 28 two example entries (out of totally 7)
include $($$g$, 28, $<$t1$>$$)$, ($b$, 28, $<$t1$>$).
Here we do not include the detailed information on which places
contain a specific word for a trajectory. It is also noteworthy that
empty cells are not indexed.
%that entry ($b$, 28, $<$t1$>$) and ($b$, 52, $<$t1, t2$>$) are next
%to each other since the cells between 28 and 52 are empty.
\end{example}

However, the above design will be problematic at query time when a
trajectory spans multiple cells, and individual fragment does not
contain all the query words, but several fragments together match
all the query words (which will become clear in the next section).
A simple fix is to associate each cell with all the words of a
trajectory. However, this significantly increases the space cost.

We next present a carefully designed mechanism that needs less space
while returning the correct results.
Suppose that a trajectory $\mathcal{TR}$ falls in $m$ cells. We
denote a trajectory fragment as $\mathcal{TR}_i$, $i \in [1, m]$,
where $\mathcal{TR}_i$ is adjacent to $\mathcal{TR}_{i+1}$ in the
trajectory. We associate words for each trajectory fragment as
follows.

\begin{itemize}
\item If $i$ is odd, the set of words for fragment $\mathcal{TR}_i$ is
the union of words in the places in the fragment.

\item If $i$ is even, the set of words for the fragment $\mathcal{TR}_i$ will be $\cup_{j=1}^{\min\{i+1,
m\}}\mathcal{TR}_j.\psi$, where $\mathcal{TR}_j.\psi$ is the union
of words in the places in fragment $\mathcal{TR}_j$.
\end{itemize}

%we associate every other segment of a trajectory with all the words
%of a trajectory.
%

\begin{figure}[!t]
 \begin{center}
  \includegraphics[width=0.6\columnwidth]{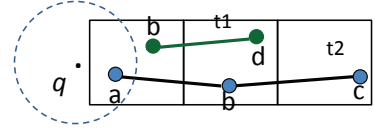}
    %\vspace{-2ex}
  \caption{Associating words with trajectory segments}\label{fig:word}
  \end{center}
  \vspace{-3ex}
\end{figure}

For example, consider trajectory $t2$ in Figure~\ref{fig:word} with
three segments in three cells. Each of the three segments contains a
term. According to the proposed mechanism, we associate segment 1
with $a$, segment 2 with $a,b,c$ and segment 3 with $c$.

This method can guarantee the correctness of the proposed algorithms
and we prove this in Lemma~\ref{lemma:fragmentword}. Note that one
cell can contain multiple fragments of the same trajectory, and the
aforementioned method is equally applicable.

% Hua:
% Here we need to explicitly say what is used to be the keys of the B+-tree.
% gao add..
\emph{Component 2}:
We use a B$^+$-tree to organize the place information and the
associated keywords in all the trajectories.
The text description for a place can be either short or long.
We use inverted list for each trajectory. Each entry consists of
three elements: \emph{trajectory ID, word ID, list of place IDs in
the trajectory}, where trajectory ID and word ID compose the key of
the B$^+$-tree.
Note that the inverted file is the most efficient index for text
information retrieval~\cite{Zobel}.

We discuss the updating process of the \cellbtree in the Appendix.

%The location information of trajectories are also organized into a
%B+tree, where each entry consists of two elements: \emph{trajectory
%ID, coordinates of places}.

\begin{algorithm}[!hbt] \small
\caption{\textsf{IE}(query $q$, result size $k$)}\label{alg:iea}

$V$ $\leftarrow$ new min-priority queue of $k$ element of $\infty$; \tcp{maintain top-k trajectories}%

$i \leftarrow 0$\;

\While{true}
{

    $rq_i$ $\leftarrow$ compute a range radius\;
    $R_i$ $\leftarrow$  construct a range with $q$ as the center and $rq_i$ as extension\;%

    \If {$i$ $\neq$0}
    {
        $R_i$ $\leftarrow$ $R_i$ $-$ $R_{i-1}$\;
    }

    %Compute the start point sp and end point ep for range $R_i$\;
    $A$ $\leftarrow$ \textsf{CTR}( $q$, $R_i$) \tcp*[r]{ See Procedure~\ref{alg:w2}}

    \For{each  trajectory $t$ in  $A$} {
        read post lists of $t$ for keywords in $q$\;

        $dist$ $\leftarrow$ \textsf{Match}($q, t, V[k]$)\tcp*[r]{ See Section~\ref{sec:alg:match}}

        \lIf{$V[k] >$ $dist$}{      $V$.\textsf{add}($dist, t$)\;}
    }

    \lIf{$V[k] <$ $rq_i$}
    {
    {\bf break}\;
    }

    $i \leftarrow i+1$\;

} {\bf return} top-$k$ trajectories in $V$ \tcp*[r]{top-$k$ results}%
%\end{algorithmic}
\end{algorithm}

\smallskip
%\noindent {\bf Remark:}
The proposed index solution \cellbtree can be implemented using
DBMSs that support the B$^+$-tree, and is update friendly. It
enables designing algorithms for processing TkSK queries that are
able to prune the search space using both types of information. In
addition to the TkSK queries, it also support other types of queries
containing a keyword component and a spatial component, e.g.,
finding trajectory containing a set of keywords within a region.
Different from the IRT baseline that takes each trajectory as a
whole and associates text information with the trajectory, in
the proposed index we use cells to divide trajectories into
segments and design an effective mechanism to associate
keywords to segments. For example, in Figure~\ref{fig:word},
given a top-1 query $q$ with keyword $b$, $t1$ is the answer.
If we use the proposed word association mechanism, we can prune
$t2$ since the segment in the first cell is associated with
word $a$ only. However, in the IRT, we cannot prune $t2$ since
it takes the trajectory as a whole.
As another example, if $q$ is to find trajectories whose segment
contains $b$ and falls in the circle, we can prune $t2$ while $t2$
cannot be pruned if treated as a whole.

%==========================================
\subsection{Incremental Expansion Algorithm (IE)}~\label{sec:alg:incremental}
%==========================================
%
We compute top-$k$ trajectories by iteratively performing range
queries with an incrementally expanded search region  on the
\cellbtree until the top-$k$ matching trajectories are retrieved.
The Incremental Expansion algorithm (\textsf{IE}) is outlined in
Algorithm~\ref{alg:iea}.
The algorithm \textsf{IE} first initializes a priority queue to
maintain top-$k$ results. We construct a range query with $q$ as the
center and a query dependent $rq_0$ as the extension (lines~4-5).

%To compute extension $rq_1$, we regard each trajectory as an object
%and estimate the distance between the query and its $k$'th nearest
%neighbor by
%
%$$D_k = \frac{2}{\sqrt{\pi}}\left[1-\sqrt{1-(\frac{k}{N)})^{\frac{1}{2}}}\right],$$
%
%where $N$ is the number of objects~\cite{Tao:2004}.
%

To compute extension $rq_0$, we take into account both keyword
information and spatial information.
% since some query keywords are more
%frequent than others.
%
Let $p(q.\psi)$ be the probability of containing $q.\psi$ as the
keyword set of a trajectory of $\mathcal{D}$. We estimate the
probability by $p(q.\psi) = \prod_{w \in q.\psi}p(w)$, where $p(w)$
can be estimated using the maximum likelihood estimation, {\textit
i.e.}, the probability of a trajectory in dataset $\mathcal{D}$ that
contains the word. We compute $rq_0$ by
$$rq_0 = \sqrt\frac{k\times L}{{\pi}\times |\mathcal{D}|\times p(q.\psi) },$$
where $L$ is the area size of the whole region, since a region of
area size $\pi\times rq_0^2$ would probabilistically contain segments of $k$
trajectories that contain all query keywords if the trajectories are
uniformly distributed in the whole region.

%By substituting $N$ with $|\mathcal{D}|.p(q.\psi)$ in the estimation
%model, we use the estimated distance $D_k$ as $rq_1$ to expand query
%region.

%
For a trajectory in the range, we  check if it contains all the
query keywords, and compute its matching distance by invoking
function \textsf{CTR} (Candidate Trajectory Retrieval, to be presented shortly).
For each returned candidate trajectory $t$ in $A$, the algorithm
invokes algorithm \textsf{Match}$(q, t, V[k])$ (See
section~\ref{sec:alg:match}) to compute the minimum match distance
between $q$ and $t$.
If the match distance of the $k$th result is larger than $rq_i$, it
is safe to terminate the algorithm because the algorithm has
considered all the trajectories that can possibly be in the top-$k$
results.
Otherwise, we compute a new range $rq_i$ = $rq_{i-1} + \tau$, where
$\tau$ is the side length of the smallest quad cell in the index. We
also tried other options, e.g., the average side length. However,
the performance of such options is worse in general.
We then retrieve trajectories in the region formed by radius $rq_i$,
but not included in the region formed by $rq_{i-1}$.

%if at least $k$ trajectories are returned at line 8, we use $V[k]$
%as $rq_i$ and trajectories outside the range cannot be a result;
%otherwise we use the largest match distance of the current
%trajectories in $V$ as $rq_i$.
%Following the work~\cite{Jensen:2004}, we extend the range by
%$D_k/k$.

%==========================================
%\subsubsection{Candidate Trajectory Retrieval}~\label{sec:ie:retrieve}
%==========================================
We proceed to present procedure \textsf{CTR}, which checks if a
trajectory in the given range $R$ contains all the query keywords.
\textsf{CTR} processes query keywords one by one. For the first
query keyword, we find the trajectories that contain the query
keyword, and intersect with the given query range $R$. Recall that
\cellbtree organizes the list of trajectories by word ID and then by
cell ID. This enables us to retrieve those trajectories that contain
the query keywords and fall in certain cells. For each subsequent
query keyword, we filter out trajectories that do not contain the
query word by scanning the corresponding cells.

\begin{procedure}[h] \small
\caption{CTR( $q$, $R$)}\label{alg:w2}

$A$ $\leftarrow$ new array; \tcp{maintain the trajectories to be checked}%

$I$ $\leftarrow$ Compute the intervals $\{(sid_j, eid_j)\}$ of
R;\tcp{ start cell id $sid_j$ and end cell id $eid_j$ }

\For{each keyword $w_i$ $(i =1,...,  |q.\psi|)$}{

    \For{each interval $I_j =$$(sid_j, eid_j)$ in $I$} %
    {
            $(C_j, I_j')$ $\leftarrow$ \textsf{getCellInInterval}($w_i$, $I_j$)\;

            \If{$I_j'$  $= \emptyset$}
            {
                remove $I_j$ from $I$\;
                continue\;
             }
            \Else
            {
                update $I$ with $I_j$\;

            }

            \If{$i =1$}{
                \For{each cell $c$ in $C_j$}
                {
                    add trajectories in $c$ to  $A$\;
                }
            }
            \Else
            {
                removes trajectories in $A$ that are not covered by
                any cell $c$ in $C_j$
            }
    }
    % add some here..

           %\If{$i =1$}{
           %     If{$|A|$ $<$ k }
           %     {
           %         add trajectories in $c$ to  $A$\;
           %     }
           % }

}

%\For{each  trajectory $t$ in  $A$} {
%    read post lists of $t$ for keywords in $q$\;

%    $dist$ $\leftarrow$ \textsf{Match}($q, t$)\;
%    $V$.\textsf{add}($dist, t$)\;
%}

{\bf return} $A$\;
%\end{algorithmic}
\end{procedure}

%Algorithms \ref{alg:w1}-\ref{alg:t1} read the disk page by page, and
%construct the cell ID intervals of the query region on the fly, as
%inspired by \cite{Ramsak:2000}. Compared with computing all the
%intervals in advance, this could avoid computing multiple intervals
%located in the same page.

The  candidate trajectory retrieval (CTR) algorithm is outlined in
Procedure~\ref{alg:w2}. It takes two arguments: query $q$ and the
given region $R$. It first computes the intervals of cell IDs that
are covered by the query region $R$ (line 2).
The algorithm proceeds to process each query word $w_i$ (lines
3--15). For each interval, it returns the trajectories that contain
keyword $w_i$ and fall in the interval.
Function \textsf{getCellInInterval}(.) returns in $C_j$ those cells that
contain word $w_i$ and fall in the interval $I_j$.
The function is implemented by following pointers between leaf nodes
of $B^+$-tree, and the jump technique~\cite{Ramsak:2000} is used to
optimize the implementation by jumping over pages.
If interval $I_j = (sid_j, eid_j)$ does not contain any cell
containing word $w_i$, we remove the interval from consideration
(lines~6--8).
Function~\textsf{getCellInInterval}(.) also returns a smaller
interval $I_j'$ if the interval covered by cells in $C_j$ is smaller
than $I_j$. We use $I_j$ to update the interval boundary $sid_j$ and
$eid_j$ (line~10).
For the first keyword $w_1$, we add the trajectories that contain
word $w_1$ and are in the region $R$ to the set of candidate
trajectories $A$ (lines~11--12). For each of subsequent keyword, the
algorithm filters the trajectories in $A$ that do not contain the
keyword (line~15). In the implementation, we organize trajectories
in $A$ by cell ID and filter trajectories in a cell if the cell does
not contain a query word.

We process query words in the ascending order of their frequencies,
{\it i.e.}, infrequent words are processed first. The reason is
 that infrequent keywords are more likely to prune
trajectories.

Before we prove the correctness of the proposed algorithm, we first
present a lemma.

\begin{lemma} \label{lemma:locate}
Consider query $q$ and a trajectory $\mathcal{TR}$ that falls in $m$
cells. We denote a trajectory fragment as $\mathcal{TR}_i$, $i \in
[1, m]$ and the cell containing trajectory fragment $\mathcal{TR}_i$
as $cell(\mathcal{TR}_i)$. If the minimum match of $\mathcal{TR}$
for $q$ that results in the minimum match distance follows in cells
$[c_1, c_2]$, $ 1 \leq c_1 \leq c_2 \leq m$, we have
$\minMatchDist(q, \mathcal{TR})$ $\geq$ $\max_{c_j \in[c_1,c_2]}
(minDist(q,c_j))$.

\proof Based on triangular inequality, we know that\\
$\minMatchDist(q, \mathcal{TR})$ $\geq$ $\Dist(q,
\mathcal{TR}_{L})$, where $\mathcal{TR}_{L}$ is a place in the
sub-trajectory of $\mathcal{TR}$ that is a minimum match of query
$q$. It is easy to see that $minDist(q,c_j)$ is not larger than
$\Dist(q, \mathcal{TR}_{L})$ where $L$ is in cell $c_j$. We complete
the proof.
\end{lemma}

According to Lemma~\ref{lemma:locate}, the minimum match distance of
a trajectory to the query is larger than the distance from query to
any cell that contain parts of the matching trajectory. We are now
ready to present the correctness of the {\textsf IE}.

\begin{lemma} \label{lemma:fragmentword}
The Incremental Expansion algorithm guarantees to find top-$k$
trajectories using \cellbtree that employs the method (in
Section~\ref{sec:alg:index}) of associating the words of the
trajectory with the different segments.

%\vspace*{6pt} \noindent {\bf Proof of
%Lemma~\ref{lemma:fragmentword}}
%
\proof
%We prove that (1) Algorithm IE is correct, that is all returned
%trajectories are desired answers; and that (2) the returned results
%are complete.
%
Suppose that a trajectory $\mathcal{TR}$ falls in $m$ cells. We
denote a trajectory fragment as $\mathcal{TR}_i$, $i \in [1, m]$. Two cases cover all the possibilities that
$\mathcal{TR}$ is one of the top-k results.

Case 1: if $\mathcal{TR}_i$ contains all the query words and the
cell containing $\mathcal{TR}_i$ is in the range of the match
distance of $k$th trajectory in the current result set (i.e.,
mindist(cell($\mathcal{TR}_i$)) $>$ minmatchdist($TR, q$)) the
trajectory $\mathcal{TR}$ will be retrieved and we compute its match
distance. In this case, algorithm~{\textsf IE}
 will not miss the trajectory $\mathcal{TR}$ if it is a
result. Note that if the cell containing $\mathcal{TR}_i$ is not in
the range, $\mathcal{TR}_i$ cannot be matching part of top-$k$
trajectory.

Case 2:  we next consider the case that none of the sub-trajectories
contains all the query keywords, but the trajectory contains all the
query keywords.
%If the subtrajectory from $TR_p$ to trq contain the
%matching part, at least two cells
We first consider that two subtrajectories in two adjacent cells
cover the query keywords. The method of associating keywords with
cells make sure that one of the two subtrajectories in the two cells
will be associated with at least the keywords of both
subtrajectories.
According to Lemma~\ref{lemma:locate}, the distance from query to
the cell containing a sub-trajectory associated with all keywords of
the two adjacent cells must be smaller than the match distance
between the trajectory and query. This grantees that
algorithm~{\textsf IE}
 will not miss the trajectory $\mathcal{TR}$ if it is a
result.
Similarly, when more than 2 adjacent cells together cover the query
keywords, at least one of the subtrajectories of $\mathcal{TR}$ in
these cells contain all the keywords of these sub-trajectories
according to the method of associating keywords to sub-trajectories.
%Its matching distance must be larger than distance to the cell that
%contain all the matching words according to lemma, So we will not
Thus, algorithm~{\textsf IE} will not miss the trajectory
$\mathcal{TR}$ if it is a result.

The two cases cover all the possibilities that a trajectory can be a
top-k result. Therefore, Algorithm IE is correct and complete.
\end{lemma}

%We need at least two sub-trajectories to answer the query. Since we
%associate the keywords with every other sub-trajectory, {\it i.e.},
%$\mathcal{TR}_2$, $\mathcal{TR}_4$,..., the distance from query to a
%sub-trajectory associated with all keywords must be smaller than the
%match distance between the trajectory and query. Thus we will not
%miss the trajectory if it is a result.

\vspace*{-10pt}
%==========================================
\subsubsection{Cost Analysis of IE Algorithm}~\label{sec:alg:cost}
%==========================================
\vspace*{-10pt}

First of all, it is noteworthy that our incremental expansion
algorithm (IE in Algorithm~\ref{alg:iea}) has an asymptotically equivalent effect
of a window search through the specific B$^+$-tree index that is also known
as a linear quadtree. In particular, such an equivalent query is
centered at query location $q.\lambda$ and its window size is
bounded by the place $\kw{L}_\mathit{last}$ that our algorithm
fetches as the last place on the trajectory that contributes to the
$k$-th minimum match distance.

To make the analysis clear, we assume that $k$ is 1, i.e., we
only get the top-1 trajectory with minimum match distance. Let
the distance from $q.\lambda$ to $\kw{L}_\mathit{last}$ through
the trajectory, i.e., the corresponding minimum match distance,
be $\mathit{Dist}(q.\lambda, \kw{L}_\mathit{last})$. This
matching distance can be used as the half window size in the
aforementioned equivalent window query.

The matching distance $\mathit{Dist}(q.\lambda,
\kw{L}_\mathit{last})$ is not a Euclidean distance since we work
with trajectories. Nevertheless, we use a Euclidean distance value
equal to $\mathit{Dist}(q.\lambda, \kw{L}_\mathit{last})$ as the
half window size in the equivalent window query. This justified by
the fact that any place out of the window thus determined must
result in a larger matching distance than does
$\kw{L}_\mathit{last}$. In this sense, our algorithm does not
need to visit any farther places out of the window. On the other
hand, we cannot reduce the window size to a value less than
$\mathit{Dist}(q.\lambda, \kw{L}_\mathit{last})$ because this
distance itself can be a Euclidean distance if all involved places
and $q.\lambda$ are in a same straight line.

Aboulnaga and Aref~\cite{DBLP:journals/dpd/AboulnagaA01} proposed a
cost model for window query processing in linear quadtrees. Given a
query window $W$ and a quadtree $T$, the model
estimates the query cost by recursively counting the quads that
overlap or are enclosed by $W$. This model can be employed here
to estimate the IO cost our algorithm incurs in searching for
trajectories with minimum match distance.

Next, we elaborate on how to estimate the minimum match distance
$\mathit{Dist}(q.\lambda, \kw{L}_\mathit{last})$ since it
determines the window query size. Let $K$ be the total number of
keywords in the entire space of interest, $C$ be the maximum number
of places per trajectory, and $Q$ be the number of keywords in $q$,
i.e., $Q=|q.\psi|$. Our analysis needs the information about the
trajectory places distribution in the space, as well as the keywords
distribution on all trajectory places. Both distributions can be
very complicated due to many hard-to-describe factors including
environment and humans. We hereby make two simplifying assumptions.
We assume there are $w$ keywords on average per trajectory place,
and no keyword is repeated across places within a same trajectory.

We count how many places our algorithm visits on the returned trajectory, i.e., the one with the minimum match distance
$\mathit{Dist}(q.\lambda, \\ \kw{L}_\mathit{last})$. The counting
starts at the first place where at least on required keyword in
$q.\psi$ is included, denoted as $\kw{L}_s$, and ends at place
$\kw{L}_\mathit{last}$. Note that both $\kw{L}_s$ and
$\kw{L}_\mathit{last}$ must have at least one required keyword
in $q.\psi$. We use $\hat{Pr}(i)$ to denote the probability that
$\kw{L}_\mathit{last}$ is the $i$-th place inclusively from
$\kw{L}_s$.

%%%

The probability of a single place containing the query words
$q.\psi$ is computed by
\vspace*{-5pt}
\begin{eqnarray}\label{eq1}
\hat{Pr}(1)=  pr(q \in \kw{L}.\psi) &= & \prod_{w \in q.\psi}(pr(w \in \kw{L}.\psi))  \nonumber\\
    & = & \prod_{w \in q.\psi}(1- pr(w \neq \kw{L}.\psi_i)^{|\kw{L}.\psi|})  \nonumber \\
   & = & \prod_{w \in q.\psi}(1- (1- pr(w))^{|\kw{L}.\psi|}),
\end{eqnarray}
where $q$ is the query,  $\kw{L}.\psi_i$ ($i \in [1, |\kw{L}.\psi|]$ ) is a
word in $\kw{L}.\psi$, and $pr(w)$ is the probability that a word in a
place $\kw{L}$ is the query word $w$.

We say that $i$ places ``jointly'' contain the query words if
1) the $i$ places cover the query words, 2) the first place and
the last place must contain some query keywords, and 3)none of
proper subsets of the $i$ places contain all the query words.
We denote the probability that $i$ places ``jointly'' contain
the query words by $\hat{Pr}(i)$. To compute it, we first
compute the probability that a subset of the $i$ places contain
the query words $pr(i)$ , which can be computed as we do in
Equation~\ref{eq1}, that is,
$$pr(i)=pr(q \in \cup_{j =1}^i \kw{L}_j)= \prod_{w \in q.\psi}(1-
(1- pr(w))^{i\cdot|\kw{L}.\psi|})$$

We next compute the probability that each place in a subset of the
$i$ places contains all the query words.
$$
p_1(i)=\sum_{j =1}^{i}
 \dbinom{i}{j}   \hat{Pr}(1)^j* (1- \hat{Pr}(1))^{i-j}
 $$
where $\hat{Pr}(1)^j$ is the probability that each of the individual
$j$ places contains all the query words, and $(1-
\hat{Pr}(1))^{i-j}$ is the probability that each of the other $i -j$
places does not contain all the query words.

We next compute the probability that a proper subset of the $i$ ($i
>2 $) places jointly contains the query words such that the subset
does not contain the first and the last places of the $i$ places and
none of single places contains all the query words.
$$
p_2(i)=\sum_{j =2}^{i-1}
 (\dbinom{i}{j} - \dbinom{i-2}{j-2})   \hat{Pr}(j)* (1- {Pr}(i-j))
$$
where $\hat{Pr}(j)$ is the probability that $j$ places of the $i$
places jointly contain the query words, and $(1- {Pr}(i-j))$ is the
probability that the other $i-j$ places do not contain the query
words.

Finally, we are ready to compute $\hat{Pr}(i)$.

\begin{equation}
\hat{Pr}(i) = pr(i) -p_1(i) - p_2(i)
\end{equation}
%where $\sum_{j =1}^{i-1}  \dbinom{i}{j}\hat{Pr}(j)$ is the
%probability of a proper subset of $i$ places containing the query
%words.

As an example, the probability that two places $L_1$ and $L_2$
jointly contain the query words $q.\psi$ is $\hat{Pr}(2)
=pr(2)-p_1(2)= pr(2) - 2 \hat{Pr}(1)*(1-\hat{Pr}(1))-\hat{Pr}(1)^2$.

\eat{ It is captured in the following.
$prob(i) = \left\{
  \begin{array}{ll}
    \frac{\dbinom{K-Q}{w - Q}}{\dbinom{K}{w}}, & \hbox{$i=1$;} \\
    \frac{P(Q, 2) \cdot \dbinom{K-Q}{i\cdot w - Q} \cdot \dbinom{i \cdot w-2}{w-1} \cdot \dbinom{i \cdot w-1-w}{w} \cdot \ldots \cdot \dbinom{2 \cdot w-1}{w} \cdot \dbinom{w-1}{w-1}}{\dbinom{K}{i\cdot w}}, & \hbox{$i>1$.}
  \end{array}
\right.$ }

%\begin{equation*}\label{equ:prob_i}
%    prob(i) = \frac{P^2_Q\cdot \dbinom{K-Q}{i\cdot w - Q} \cdot \dbinom{i \cdot w-2}{w-1} \cdot \dbinom{i \cdot w-1-w}{w} \cdot \ldots \cdot \dbinom{2 \cdot w-1}{w} \cdot \dbinom{w-1}{w-1}}{\dbinom{K}{i\cdot w}}
%\end{equation*}

%\begin{equation*}\label{equ:prob_i}
%prob(i) = \left\{
%  \begin{array}{ll}
%    \frac{\dbinom{K-Q}{w - Q}}{\dbinom{K}{w}}, & \hbox{$i=1$;} \\
%    \frac{P(Q, 2) \cdot \dbinom{K-Q}{i\cdot w - Q} \cdot \dbinom{i \cdot w-2}{w-1} \cdot \dbinom{i \cdot w-1-w}{w} \cdot \ldots \cdot \dbinom{2 \cdot w-1}{w} \cdot \dbinom{w-1}{w-1}}{\dbinom{K}{i\cdot w}}, & \hbox{$i>1$.}
%  \end{array}
%\right.
%\end{equation*}

Consequently, the expected number of places to visit is
$\sum^{C-1}_{i=1} i \cdot \hat{Pr}(i)$. Assuming that the average
segment length of all trajectories is $len$, the expected distance
from place $\kw{L}_s$ to place $\kw{L}_\mathit{last}$ is
$len \cdot \sum^{C-1}_{i=1} i \cdot \hat{Pr}(i)$.

Finally, we estimate the Euclidean distance between query location
$q.\lambda$ and place $\kw{L}_s$, i.e.,
$\mathrm{Dist}(q.\lambda, \kw{L}_s)$. Suppose there are $Y$
trajectories in the entire space, which results in $Y \cdot C$
places in total. The average Euclidean distance between two adjacent
places is $L / \sqrt{Y \cdot C}$, where $L$ is the side size of the
entire space. On average we need to visit $\lceil K / w \cdot
Q\rceil$ places to see a required keyword in $q.\Psi$. As a result,
$\mathrm{Dist}(q.\lambda, \kw{L}_s)$ is approximated by $L /
\sqrt{Y \cdot C} \cdot \lceil K / w \cdot Q\rceil$.

To put it altogether, $\mathit{Dist}(q.\lambda,
\kw{L}_\mathit{last}) \approx L / \sqrt{Y \cdot C} \cdot \lceil
K / w \cdot Q\rceil + len\cdot \sum^\kw{L}_{i=1} i \cdot
\hat{Pr}(i)$. As mentioned above, this distance and the query
location $q.\lambda$ together determine the window query whose cost
can be estimated using the model proposed by Aboulnaga and
Aref~\cite{DBLP:journals/dpd/AboulnagaA01}.

%\input{analysis}

%==========================================
\subsection{\!\!Computing Match Distance of a Trajectory}
\label{sec:alg:match}
%==========================================
We present algorithm \MatchDist for searching the minimum match of a
selected trajectory to a query, and computing the match distance.
\MatchDist is invoked by algorithm \textsf{IE} and our baseline
algorithms.

Given a trajectory $\mathcal{TR}$ = $\kw{L}_1, ..., \kw{L}_n$ and a
query $q$, a naive approach to finding the minimum match is to check
all possible sub-trajectories in $\mathcal{TR}$. For each sub-trajectory, we check if it is
a match of the query $q$; if it is, we compute the match distance.
Finally, we get the minimum match distance.
The time complexity of the naive approach is $O(|\mathcal{TR}|^2)$.

We proceed to develop an approach with $O(|\mathcal{TR}|)$
complexity based on the principle of divide and conquer and the idea
of dynamic programming.
Specifically, we divide the problem into sub-problems, each of which
is to search the minimum match starting from a place in a trajectory
$\mathcal{TR}$. At each place, we check whether query $q$ can be
matched by a sub-trajectory starting at the place.  Here a key idea
is that we reuse the computation of the sub-problem of finding the
minimum match sub-trajectory starting at the preceding place for
processing the sub-problem of finding matching sub-trajectory
starting at the current place.
After we process all the sub-problems, we will find a minimum match,
if any.

%Before presenting the algorithm,
We now introduce lemmas required for developing the algorithm.
Based on Definition~\ref{def:match}, we have the following
proposition.

\begin{proposition} \label{prop:match}
If a sub-trajectory $\mathcal{TR}.\kw{L}_{s}^{e}$ from place
$\kw{L}_s$ to place $\kw{L}_e$ matches $q$, then any sub-trajectory
containing $\mathcal{TR}.\kw{L}_{s}^{e}$ matches $q$. If a
sub-trajectory $\mathcal{TR}.\kw{L}_{s}^{e}$ is not a match of $q$,
then any sub-trajectory of $\mathcal{TR}.\kw{L}_{s}^{e}$ is not a
match.
\end{proposition}

\begin{lemma} \label{lemma:problem}
If a sub-trajectory $\mathcal {TR}.\kw{L}_{s}^{e}$ is a minimum
match of a query $q$,
 and sub-trajectory $\mathcal {TR}.\kw{L}_{ps}^{ed}$ is a match of query
 $q$ such that $ps \leq s$ and $ed \geq e$,
 then  matchDist(q, $\mathcal {TR}.\kw{L}_{s}^{e}$) $\leq$ matchDist(q, $\mathcal {TR}.\kw{L}_{ps}^{ed}$).

\proof
%\vspace*{6pt} \noindent {\bf Proof of Lemma~\ref{lemma:problem}}
%
We can prove the lemma by the distance triangle inequality. The
distance between $q$ and $\kw{L}_{s}$ must be smaller than the sum
of the distance between $q$ and $\kw{L}_{ps}$ and the distance
between $\kw{L}_{ps}$ and $\kw{L}_{s}$.
\end{lemma}

Based on Lemma~\ref{lemma:problem}, we have the following
proposition.

\begin{proposition}\label{lemma:nofurther}
%Corollary:
If a sub-trajectory $\mathcal{TR}_1$ is contained by sub-trajectory
$\mathcal{TR}_2$, the match distance of $\mathcal{TR}_1$ to query
$q$ is smaller than that of $\mathcal{TR}_2$.
\end{proposition}

\begin{lemma} \label{lemm:boundsub}
Let sub-trajectory $\mathcal {TR}.\kw{L}_{s}^{e}$ be a match of
query $q$. The maximal distance of all places in $\mathcal
{TR.L}_{s}^{e}$ to query $q$, i.e., \\$\max_{i \in [s, e]} dist(q,
TR.L_i)$,
is a lower bound of the match distance between the sub-trajectory and $q$.\\
\proof  The proof can be established based on triangle
inequality.
\end{lemma}

The pseudocode of the algorithm is outlined in
Procedure~\ref{alg:match}.
The algorithm takes in three arguments, a query $q$, a trajectory
$\mathcal{TR}$, and the match distance of the current $k$th result.
It uses a variable \kw{mDist} to keep track of the current minimum
match distance, and \kw{ts} and \kw{te} to track the start place and
end place, respectively, of the corresponding minimum match(line 1).
It uses an array $C$ to keep track of the number of occurrences of
query keywords (in query $q$) in a sub-trajectory (line 2). It uses
a variable $b$ to represent the start place of a sub-trajectory, and
a variable $ll$ to represent the end place of a sub-trajectory.
The algorithm initializes array $C$ with the occurrences of query
keywords in location $\kw{L}_1$ (line 3).

%\begin{minipage}{3.36in}
\begin{procedure}[!t]\small \caption{ Match( query $q$,
trajectory $\mathcal{TR}$, distance $\xi$)}\label{alg:match}
\LinesNumbered \SetKwFunction{IsMatch}{IsMatch}
\SetKwData{mDist}{mDist} \SetKwData{ts}{ts} \SetKwData{te}{te}
        $\mDist \leftarrow$ $\infty$; $\ts \leftarrow \infty$; $\te \leftarrow \infty$\tcp*[r]{\footnotesize Result variables}
        $C$ $\leftarrow$ an array of $|q.\psi|$ elements of 0 \tcp*[r]{\footnotesize used as the $~~~~~~~~~~~~~~~~~~~~~~~$counter for each query word}

            \lFor{ each word $w$ in $\kw{L}_1.\psi$}
            {
                $C[w]$ $\leftarrow$ $C[w]+1$\;
            }
%        \State $ll$ $\leftarrow$ $1$ \Comment{id of the last scanned location}

        $ll$ $\leftarrow$ $1$   \tcp*[r]{\footnotesize the last scanned place}
        $b$ $\leftarrow$ $1$\;
%        \For {each place $\kw{L}_{b}$ in the order of $<$$\mathcal{TR}.\kw{L}_1$,...,$\mathcal{TR}.\kw{L}_n$$>$}
        \While{$ll \leq n$}
            {
            ($ism$, \mDist, \ts, \te) $\leftarrow$ \IsMatch($q$,  $C$,   $b$, $ll$, \mDist)\;%
            \If{$ism$}
            {
               \lFor {each word $w$ in $\kw{L}_b.\psi$}
                {
                     $C[w]$ $\leftarrow$ $C[w] -1$\;
                 }
            $b$ $\leftarrow$ $b +1$; {\bf continue}\;}
            %\While{$ll < n$}
            %{
                $ll$ $\leftarrow ll+1$\;

                \If{${\rm Dist}(q, \kw{L}_{ll}) > \xi$ }
                {
                    $b$ $\leftarrow$ $ll$ +1\;
                    $C$ $\leftarrow$  0\tcp*[r]{\footnotesize for all elements of \!\!\!\!\!$C$~~}

                     {\bf continue}\;
                }

                \For{ each word $w$ in $\kw{L}_{ll}.\psi$}
                {
                    $C[w]$ $\leftarrow$ $C[w]+1$\;
                }

                \If{$\min({\rm Dist}(q, \kw{L}_{b}), {\rm Dist}(q,\kw{L}_{ll}))$ + $\sum_{j=b}^{ll-1} {\rm Dist}(\kw{L}_j,\kw{L}_{j+1})$ $> \xi$}
                {
                    \lFor {each word $w$ in $\kw{L}_b.\psi$}
                    {
                        $C[w]$ $\leftarrow$ $C[w] -1$\;
                    }
                    $b$ $\leftarrow$ $b +1$; {\bf continue}\;
                }

                ($ism$, \mDist, \ts, \te) $\leftarrow$ \IsMatch($q$,  $C$, $\mDist$, $b$, $ll$)\;%
                \If{$ism$}{
               \lFor {each word $w$ in $\kw{L}_b.\psi$}
                {
                     $C[w]$ $\leftarrow$ $C[w] -1$\;
                 }
                $b$ $\leftarrow$ $b +1$\;}

            %}
            \If{$ll$ = $n$ and {\bf not}($\forall$ $w \in q.\psi$, $c[w]$ $>$ 0)}
            {
                {\bf break} \tcp*{\footnotesize no remaining matches}
            }
            %\State $ll$ $\leftarrow$ $k$
        }
     {\bf return} ($\mDist, \ts, \te$)\; %\Comment{top-$k$ results }
%\label{alg:matchdistance}
\end{procedure}
%\end{minipage}
\begin{procedure}[!t]
\SetKwFunction{IsMatch}{IsMatch} \SetKwData{mDist}{mDist}
\SetKwData{ts}{ts} \SetKwData{te}{te}
 \small \caption{{IsMatch}( $q$,
$C$,  $\mathsf{ts}$, $\mathsf{te}$,
$\mathsf{mDist}$)}\label{alg:ismatch}
\KwIn{query $q$, counter vector $C$, start place $\ts$, end place
$\te$, match distance $\mDist$)}
\KwResult{$ism$, $\mDist$, $\ts$ , $\te$ }
\BlankLine
           \If(\tcp*[f]{\footnotesize it is a match}) {$\forall$ $w \in q.\psi \;$ $C[w]$ $>$ 0}
            {
                $md$ = $\min({\rm Dist}(Q, \kw{L}_{\ts}), {\rm Dist}(Q, \kw{L}_{\te}))$ + $\sum_{j=\ts}^{\te-1} {\rm Dist}(\kw{L}_j, \kw{L}_{j+1})$\;%
                \lIf {$\mDist > md$}
                {
                    $\mDist \leftarrow$ $md$\; %$s \leftarrow b$; $e
                    %\leftarrow                     ll$\;
                }
                % remove count of the first
%                \For {each word $w$ in $\kw{L}_b.\psi$}
%                {
%                     $C[w]$ $\leftarrow$ $C[w] -1$\;
%                 }
                 %{\bf continue} the for-loop\;
                 return ($true$, \mDist, \ts, \te)\;
            }
            return ($false$)\;
\end{procedure}

For each place $\kw{L}_{ll}$, Procedure \MatchDist searches for a
match for sub-trajectories from $\kw{L}_b$ to $\kw{L}_{ll}$ (lines
7--24).
% until the last location $\kw{L}_n$.
%
Procedure \MatchDist scans places to the right of $\kw{L}_b$ to see
whether a sub-trajectory starting from $\kw{L}_b$ exists to match
query $q$ (lines 7--17).
During the scan, \MatchDist updates the counter for each query
keyword when it encounters a new place $\kw{L}_{ll}$ (lines 9--10).

Based on the minimum match distance $\xi$ of the current $k$th
result, we develop two punning strategies.

{\it Pruning 1:}~If we encounter a place $\kw{L}_{ll}$ (line 12)
such that the distance  {\sf Dist}$(L_{ll}, q)$ is larger than the
minimum match distance $\xi$ of the current $k$th result, any
sub-trajectory containing $\kw{L}_{ll}$ cannot be a result according
to Lemma~\ref{lemm:boundsub}. Hence, any sub-trajectory starting
from a place between current $L_b$ and $L_{ll}$ cannot be a top-$k$
result and we will skip to the next point $L_{ll+1}$ to search
sub-trajectory starting from $L_{ll+1}$ (line~15), which is
equivalent to invoke procedure \MatchDist to find the minimum
matching for sub-trajectory starting from $L_{ll+1}$.

{\it Pruning 2:}~If $\min({\rm Dist}(q, \kw{L}_{b}), {\rm Dist}(q,
\kw{L}_{ll}))$ + $\sum_{j=b}^{ll-1} {\rm Dist}(\kw{L}_j,
\kw{L}_{j+1})$ is larger than the match distance $\xi$,
sub-trajectories starting from $b$ to the right cannot be a top-$k$
result (due to triangle inequality). Hence, we move to the next
start place $\kw{L}_{b+1}$ (line~20).

The algorithm invokes procedure \IsMatch (to be explained shortly)
to check whether a sub-trajectory starting from $\kw{L}_b$ and
ending at $\kw{L}_{ll}$ is a match of query $q$ and to compute the
match distance for a match (lines 7 and 21).

{\it Pruning 3:}~If we find a match, we stop scanning further to the
right (lines 10 and 24).
% for other
%sub-trajectories starting from $\kw{L}_b$.
This is because the sub-trajectories generated by further scanning
contain the match sub-trajectory from $\kw{L}_b$ to $\kw{L}_{ll}$,
and thus will have larger match distance than that of the current
one according to Proposition~\ref{lemma:nofurther}.

If we find a match, we eliminate the contribution of place
$\kw{L}_b$ from $C$ by reducing the counter $C[w]$ by 1 if word $w$
appears in $\kw{L}_b.\psi$ (lines 9 and 23).
%
%Note that $ll$ records the place in $\mathcal{TR}$ that has been
%lastly scanned in the previous iteration (corresponding to
%$\kw{L}_{b}$).
%
After the elimination, $C$ only records the frequencies of query
keywords in sub-trajectory from $\kw{L}_{b+1}$ to $\kw{L}_{ll}$.
%
%
%, and $C$ will be reused in the next start place $\kw{L}_{b+1}$.
%
This enables us to reuse the computation at $\kw{L}_b$ to search
matching sub-trajectory starting at $\kw{L}_{b+1}$. Any
sub-trajectory between $\kw{L}_{b+1}$ and $\kw{L}_{ll-1}$ must not
be a match since they are contained by the sub-trajectory from
$\kw{L}_{b}$ to $\kw{L}_{ll-1}$, which is not a match.
Hence, to find match sub-trajectory starting from $\kw{L}_{b+1}$, we
do not need to check these sub-trajectories. Instead,
we check the sub-trajectories starting from $\kw{L}_{b+1}$ and
ending at $\kw{L}_{ll}$ (line 7), and beyond if required (lines
11-24).
%
%To do this, we reuse the computation from the previous iteration---
%before line 10 is executed the counter $C$ keeps the frequency of
%query words from $\kw{L}_{b+1}$ to $\kw{L}_{ll-1}$.
%
%
In other words, we only need to scan from location $ll$, rather than
the start location $\kw{L}_{b+1}$, due to reusing the computation at
$\kw{L}_b$.

{\it Pruning 4:}~If the sub-trajectory from $\kw{L}_b$ to the last
place $\kw{L}_n$ cannot match query $q$, the algorithm terminates
(lines 25--26)
%
%If trajectory $TR$ does not match $Q$, we conclude that
since any sub-trajectory of the sub-trajectory from $\kw{L}_b$ to
$\kw{L}_n$ cannot match $q$ according to
Proposition~\ref{prop:match}.

%The algorithm begins with the first place $\kw{L}_1$ in $TR$.
We proceed to present Procedure \IsMatch. If every query keyword in
$q$ is included in the sub-trajectory from \kw{ts} to \kw{te} (line
1), the sub-trajectory matches query $q$, and the Procedure computes
the match distance $md$ (line 2), and updates \kw{mDist} with
$md$(line 3).

The correctness of the algorithm is obvious: If there exists a
minimum match in $TR$ for query $q$, the match must starts with a
place in $TR$, our algorithm is able to find the minimum match
starting from each place, and thus is able to find a minimum match
if there is one.

%\begin{example}
%add an example here {\bf Meihui, could you add one?}
%\end{example}

%We extend \MatchDist by two pruning strategies to compute the
%minimum match distance between a trajectory $\mathcal{TR}$ and query
%$q$.
%
%We proceed to present the pruning strategies.
%\emph{The pruning strategy (to be detailed}):

\smallskip

\noindent{\bf Complexity}: Procedure \MatchDist is a linear time
algorithm, and its complexity is $O(|\mathcal{TR}|)$. Note that the
words of each location are processed twice at most (once as the end
of a sub-trajectory and the other as the head).
%
%the while-loop (line 10), the value of $ll$ is inherited from the
%last loop. Actually, each place is considered at most once in the
%for loop and once in the while loop.
Two tricks in procedure
\MatchDist are essential to achieve the linear complexity: 1) we
divide the task to sub-problems of finding the minimum match
starting from each place; and 2) we are able to reuse the
computation for the sub-problem in the preceding place.

%==========================================
\section{Experiments}\label{sec:exp}
%==========================================

We conduct extensive experiments on real trajectory datasets to
study the performance of the proposed index \cellbtree for answering
T$k$SK queries. We build our proposed index in BerkeleyDB.
%, which is widely
%used for key/value data storage and retrieval.
In the following
experiments, our approach is compared with four baseline algorithms,
including IF, RT, IRT and ITB-tree. ITB-tree is presented in
Appendix~\ref{ssec:itb-tree}.

\subsection{Experimental Settings}\label{sec.settings}
We crawl three real spatial trajectory datasets, located in US,
France and Germany, respectively, from online travel route sharing
web sites \footnote{www.bikely.com}$^,$\footnote{www.gpsies.com}. In
the US dataset, there are $12,832$ trajectories and each trajectory
contains around $60$ locations. France dataset contains $27,689$
trajectories and each trajectory contains around $78$ locations. The
Germany dataset %is the biggest one. We sample it to a reasonable
contains $40,000$ trajectories and each trajectory contains an
average of $40$ locations. We use a real question and answer dataset
to attach text to the locations in each trajectory. The dataset is
publicly available from Yahoo! Webscope and contains $3,895,298$
questions and their answers (Q\&As), written in English. Dataset
France and Germany are generated by randomly selecting a question
for a location in the France and Germany trajectories. For the US
dataset, we attach both a question and its answer to the locations
in the US data. That is, the trajectories in the US dataset are
associated with much more keywords than those in the other two
datasets.

In addition to the data from online route sharing web sites, we also
generate a real trajectory dataset from Flickr. We retrieve photos
in New York City with shotting time, geo-location and descriptive
tags from the same user and used them to generate trajectories based
on the approach~\cite{Lu:2010}. This dataset contains $19,104$
trajectories and each trajectory contains around $4$ locations.

%However, its number of distinct locations is the smallest among the
%four data sets.

The detailed statistics of the three generated datasets are given in
Table~\ref{tbl:datasets}.

\begin{table}[htp]
\centering
\begin{tabular}{|l|c|c|c|c|}
\hline
 & {\bf US} & {\bf France} & {\bf Germany} & {\bf Flickr} \\
\hline
\#traj & $12,832$ & $27,689$ & $40,000$ & $19,104$\\
\hline
\#location & $760,516$ & $1,608,412$ & $1,314,243$ & $55,059$ \\
\hline
\#word & $26,792,407$ & $9,098,284$ & $5,620,720$ & $2,654,477$ \\
\hline
\#{\small distinct-word} & $452,734$ & $244,779$ & $164,882$ & $58,917$ \\
\hline
\end{tabular}
\caption{Datasets statistics} \label{tbl:datasets}
\end{table}

In order to evaluate the scalability, we also generate datasets with
different number of trajectories and different number of locations
per trajectory by sampling the Germany dataset. The number of
trajectories increases from $10$K to $40$K and the number of
locations in each trajectory increases from $50$ to $200$
respectively. We list the settings in Table~\ref{tbl:parameters},
where the default values are shown in bold.

\begin{table}[htp]
\centering
\begin{tabular} {|l|l|}
\hline
{\bf Parameter} & {\bf Setting}\\
\hline
Datasets & US, FR, GM, Flickr \\
%\# of trajectories & $10$K, \textbf{20K}, $30$K, $40$K \\
%\# of locations per trajectory & {\bf $50$}, $100$, $150$, $200$ \\
\hline
\# of queries & {\bf 50} \\
$k$ in T$k$SK query & \textbf{5}, $10$, $15$, $20$, $25$ \\
\# of keywords in T$k$SK query & $2$, {\bf 3}, $4$, $5$ \\
\hline
\# of segments per quad cell & $400$, $600$, {\bf 800}, $1000$, $1200$\\
\hline
\end{tabular}
\caption{Experimental parameters and settings}
\label{tbl:parameters}
\end{table}

As shown in Table~\ref{tbl:parameters}, for each of the dataset we
randomly generate a set of $50$ queries and we report the average
running time. I/O cost is not reported in the experiments because
inverted file, R-tree and BerkeleyDB have different file I/O
mechanisms and it is difficult to find an appropriate and fair
comparison method in terms of I/O cost. In the experiments, we vary
the number $k$ in the T$k$SK query from $5$ to $25$. To study the
effect of the number of query keywords, we vary it from $2$ to $5$.
Recall that our indexing approach relies on a grid partitioning of
the spatial spaces. We also investigate the performance implications
of different partitioning granularities. In particular, we vary the
number limit of trajectory segments per cell from 400 to 1200. All
the algorithms including the baselines are implemented in Java and
run on a server installed with Centos operating system.

\subsection{Query Performance}\label{sec.performance}
\subsubsection{Effect of $k$ in T$k$SK queries}
\begin{figure*}[!th]
\centering
\begin{tabular}{c}
\hspace{-14pt} \subfigure[Running time on US]{
\psfig{figure=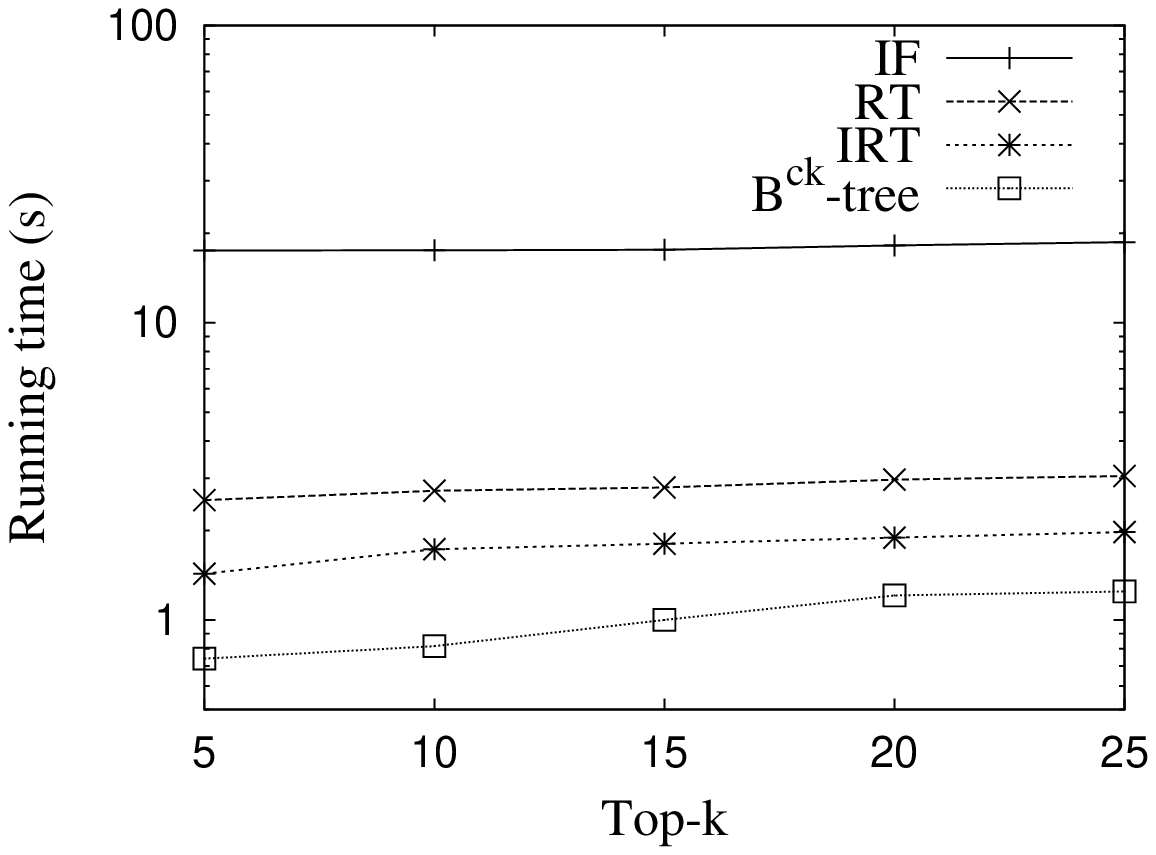,width=0.38\linewidth}
}\hspace{0.8in}\subfigure[Running time on GM]{ \vspace{2pt}
\psfig{figure=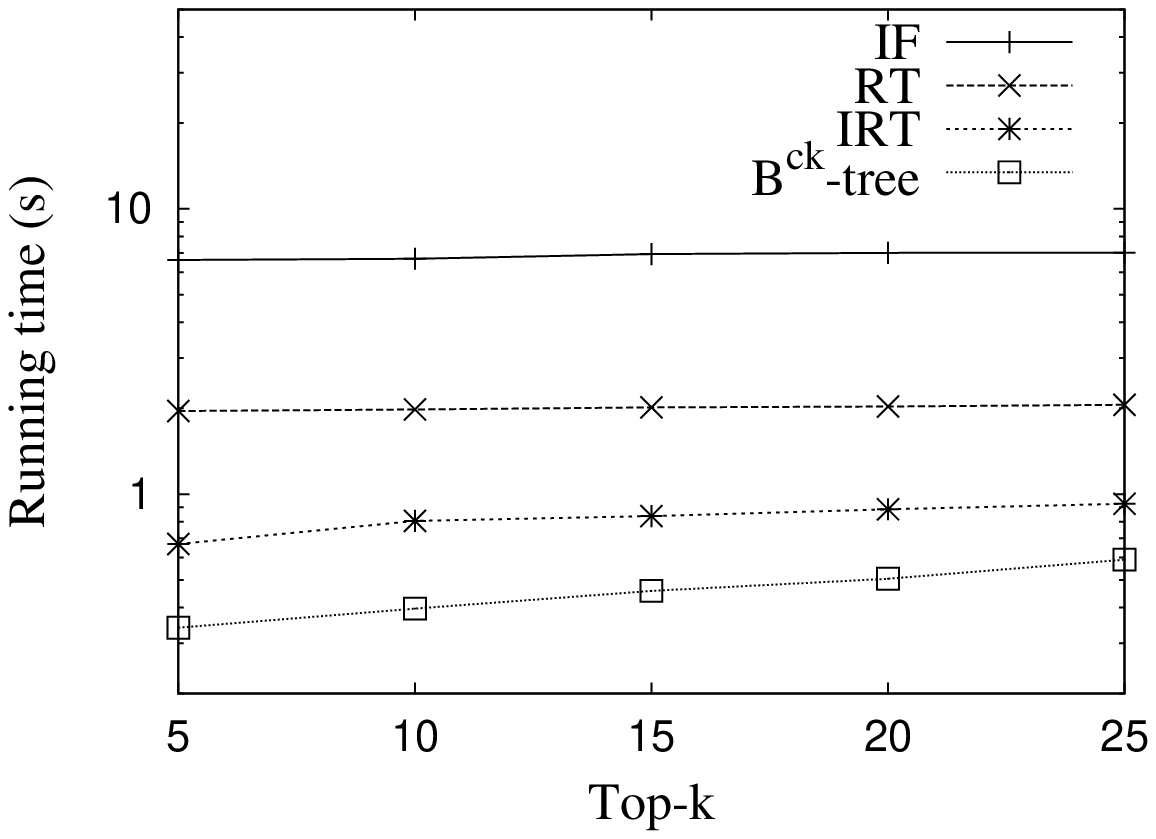,width=0.38\linewidth} }

\\

\hspace{-14pt} \subfigure[Running time on FR]{
\psfig{figure=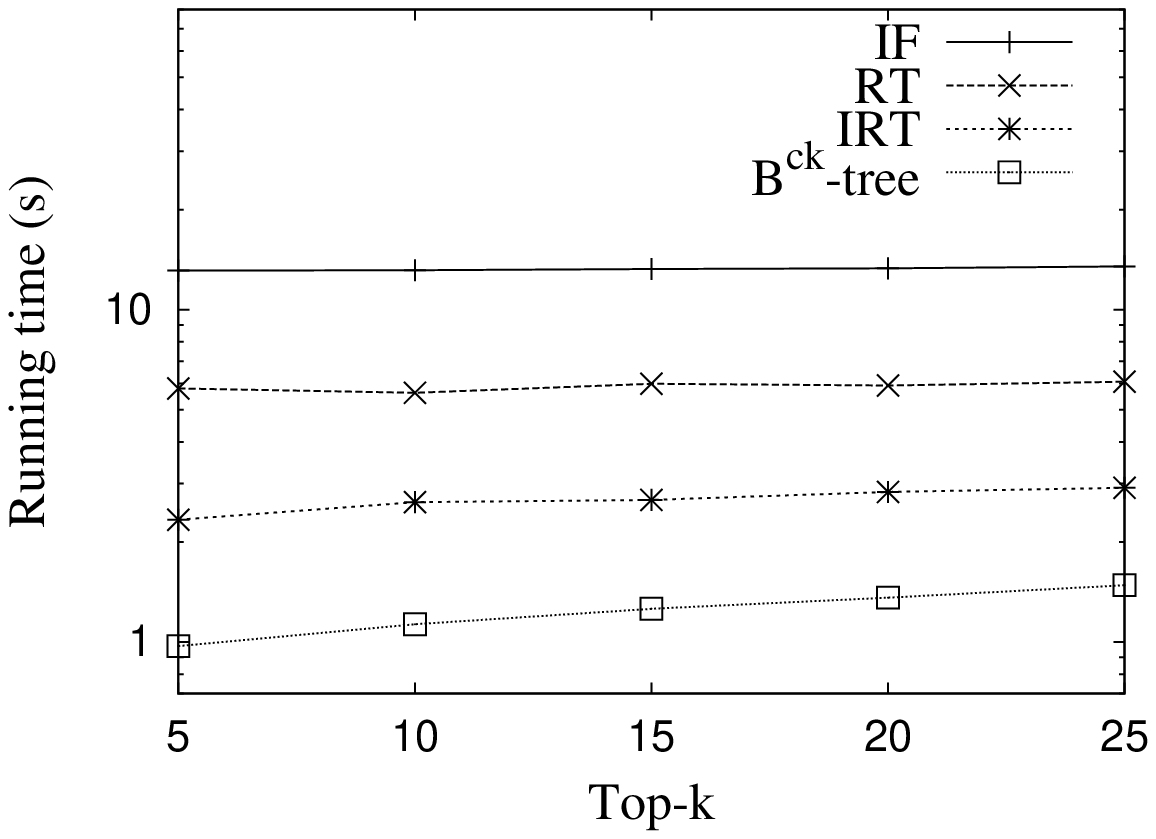,width=0.38\linewidth}
}\hspace{0.8in}\subfigure[Running time on Flickr]{
\psfig{figure=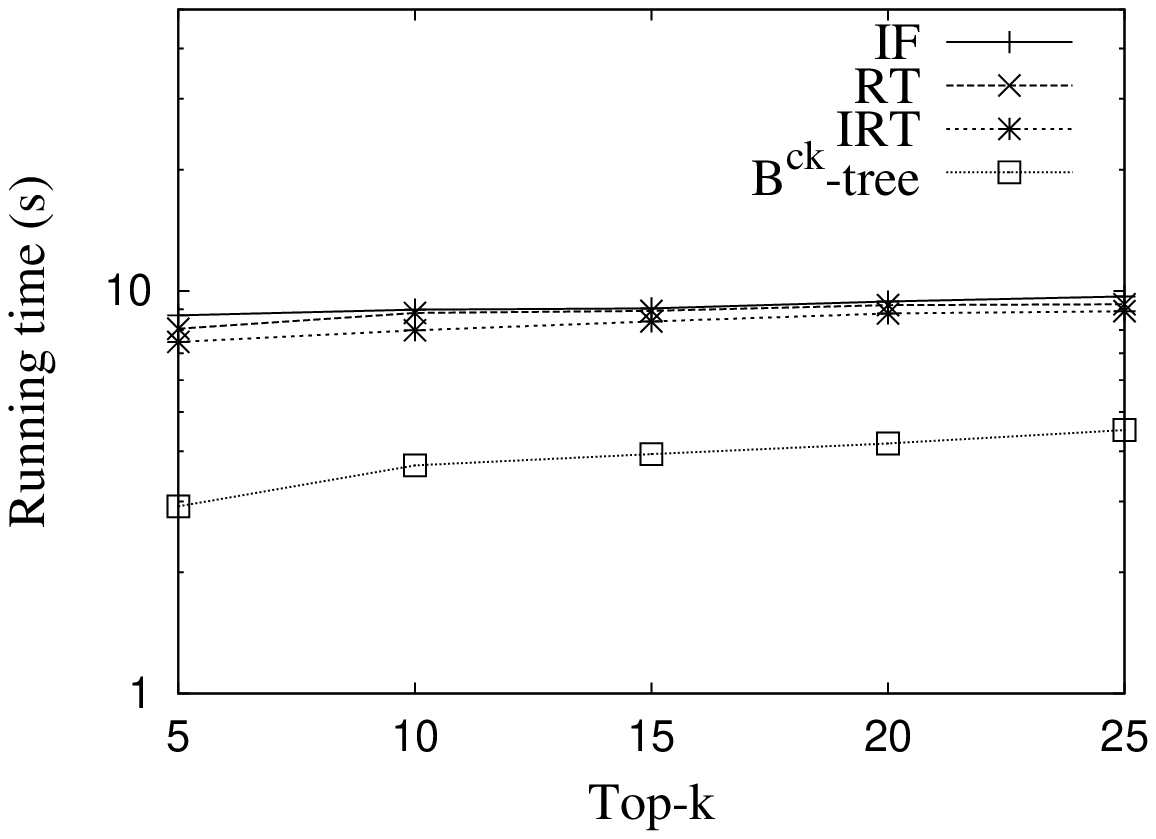,width=0.38\linewidth} }

\end{tabular}
\caption{Varying $k$ in T$k$SK queries} \label{fig:topk}
\end{figure*}

In the first set of experiments, we fix the number of query keywords
at $3$ and study the effect of $k$ in the top-$k$ queries. We plot
the average running time on the four real datasets in
Figure~\ref{fig:topk}. We notice that ITB incurs much higher cost
than the other indexes. For instance, in the US dataset, the running
time of ITB is about $3$-$6$ times higher than IF and more than $10$
times higher than our approach using \cellbtree. ITB's relatively
low performance is attributed to two reasons. First, ITB indexes
locations rather than trajectories. Second, a leaf node in ITB only
contains the locations from the same trajectory. Hence, the ITB
index contains much more nodes than do the other indexes. In order
to make the figures more presentable, we do not present the results
of ITB in the figures in this section.

Figure~\ref{fig:topk} shows that our indexing approach significantly
outperforms the other three baseline approaches in all datasets.
Note that y-axes are in logarithmic scale.
\cellbtree is usually around 1-2 times faster than IRT, the best
baseline among the four baselines.
Since IF finds all the trajectories that match the query, the
running time remains constant for all values of $k$. The other three
methods, on the other hand, incur higher cost as $k$ increases. This
is expected since they use the match distance of the $k$th
trajectory as the pruning condition. We observe that IRT performs
better than RT on datasets US, GM and FR while IRT and RT perform
almost the same on dataset Flickr.
IRT uses the IR-tree ~\cite{vldb09} to prune search space utilizing
both spacial information and text information. IRT is effective on
US, GM and FR, in which trajectories are distributed over a whole
country, and thus the overlap among the MBRs of trajectories is
relatively small although IRT takes a whole trajectory as an object.
On the three datasets, RT is worse than IRT since RT is based on the
R-tree and only uses spatial information to prune search space.
However,  trajectory data from Flickr is from a city, and
 simply treating a whole trajectory as an object
yields very high overlap between MBRs and thus degrades the pruning
power of text information of the IR-tree used in the IRT algorithm.
The overlap between MBRs also explains why RT performs poor on
dataset Flickr.

\subsubsection{Effect of the number of query keywords}
\begin{figure}[!th]
\centering
\begin{tabular}{c}
\hspace{-14pt} \subfigure[US]{
\psfig{figure=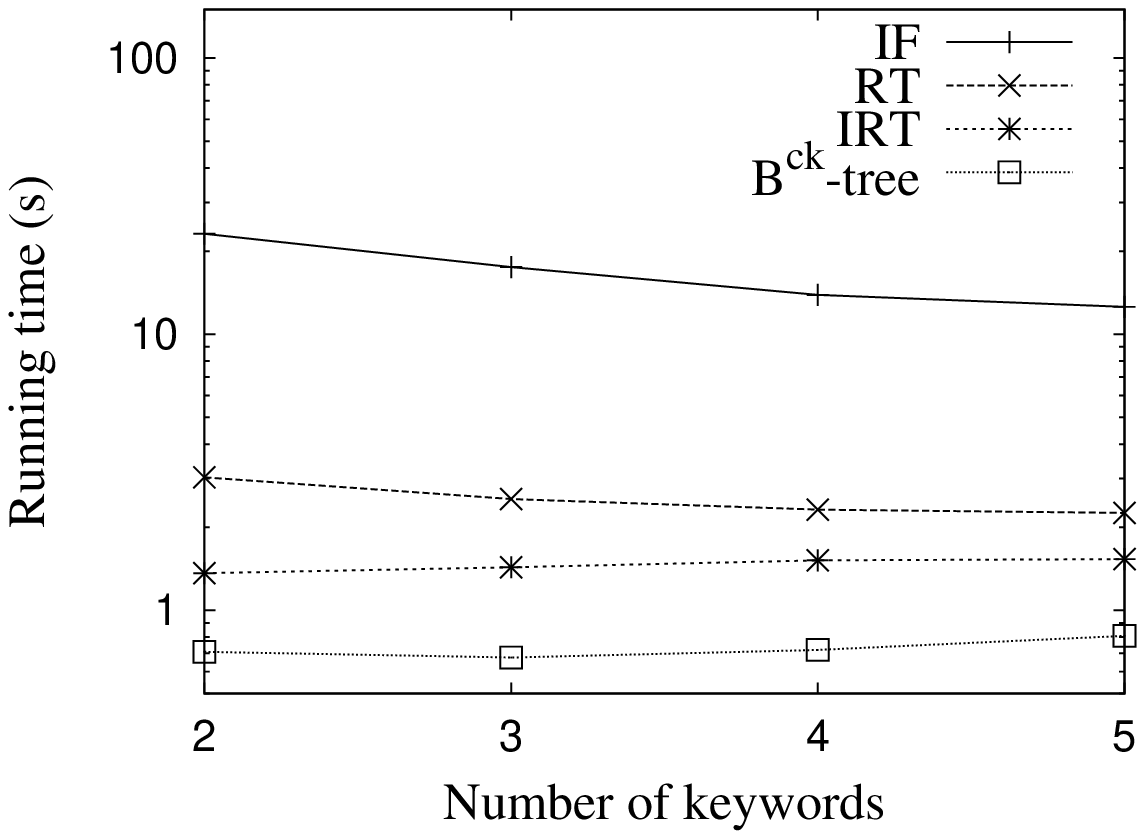,width=0.5\linewidth} }
\subfigure[Germany]{
\psfig{figure=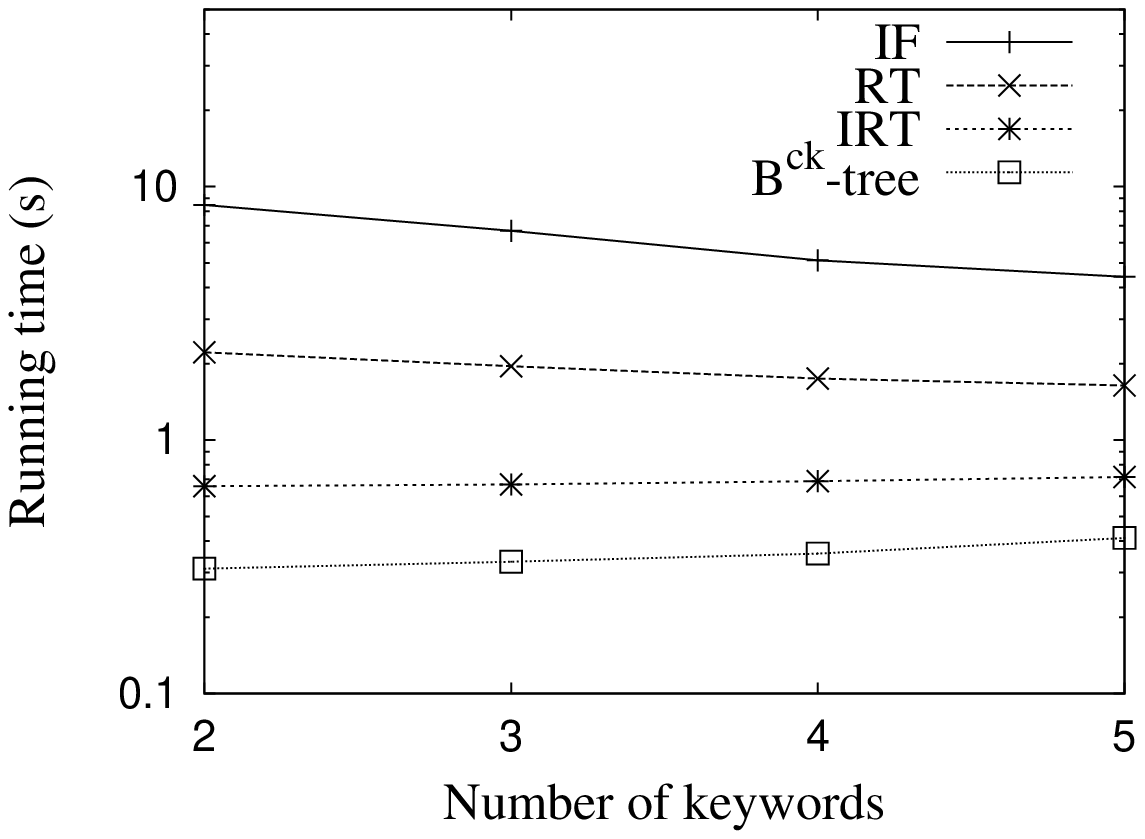,width=0.5\linewidth} }

\\

\subfigure[France]{ \hspace{-14pt}
\psfig{figure=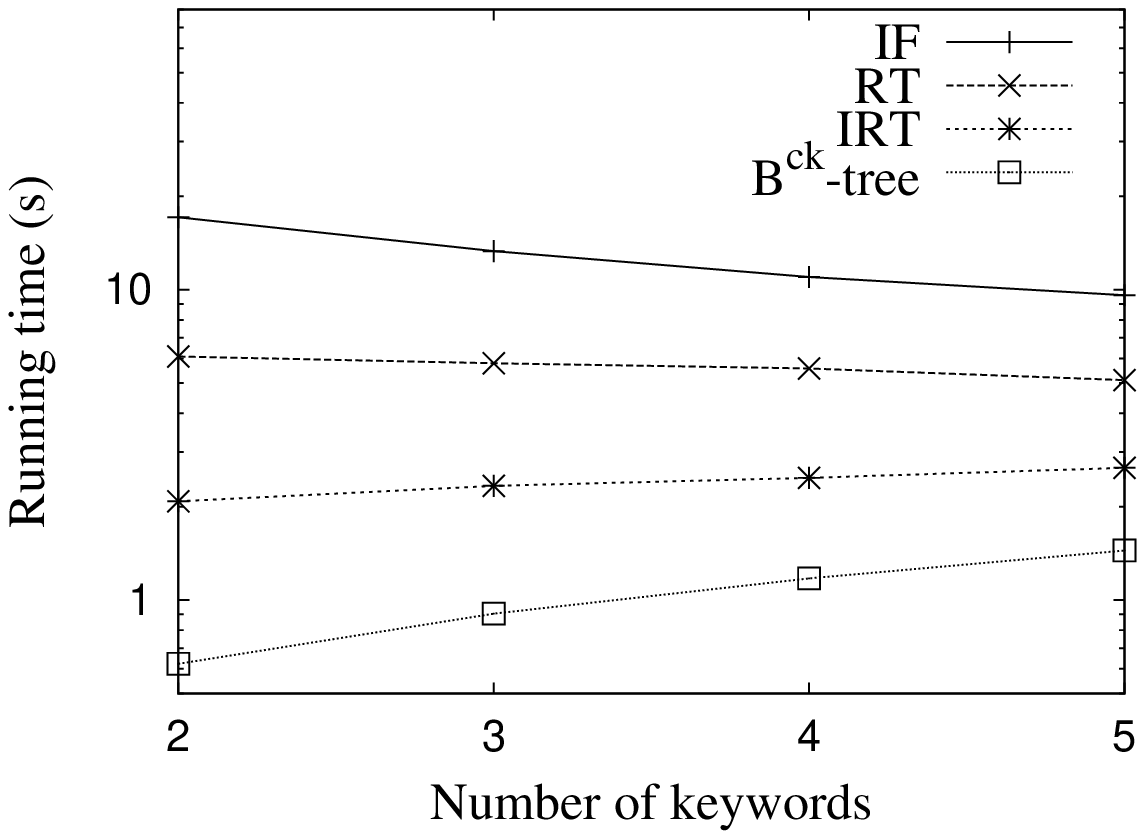,width=0.5\linewidth} }
\subfigure[Flickr]{
\psfig{figure=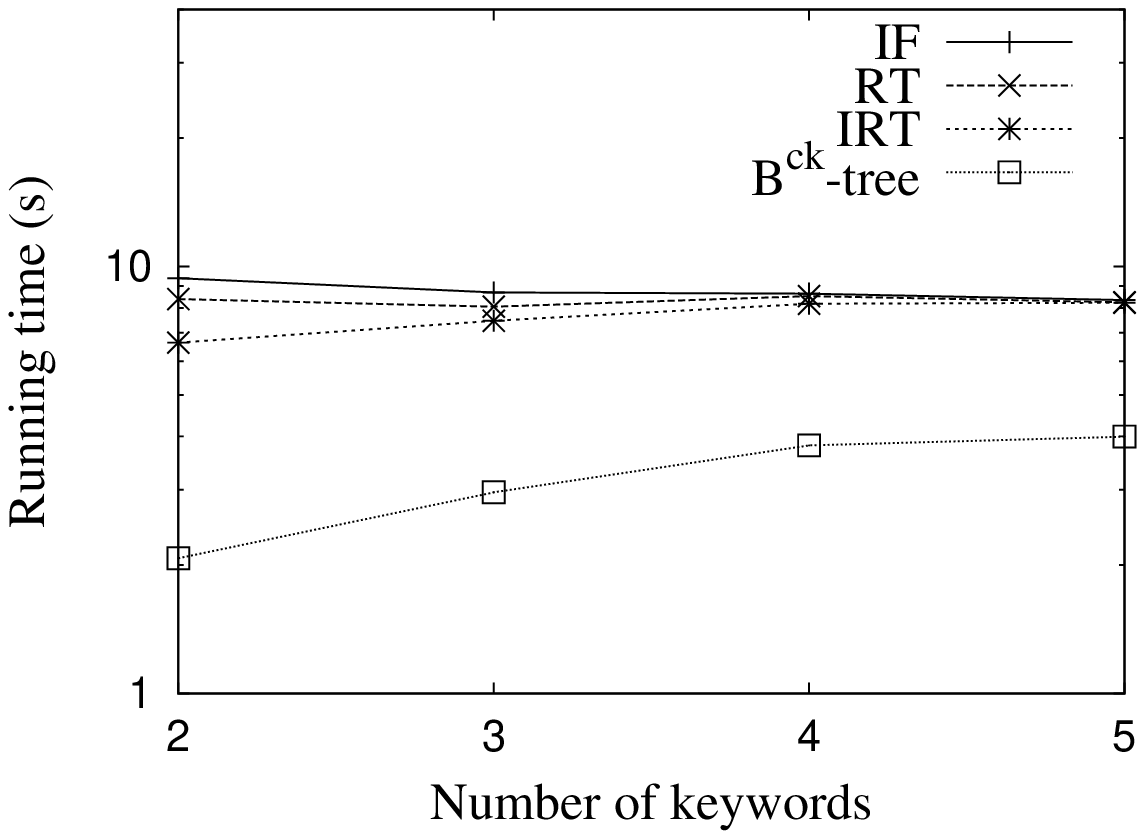,width=0.5\linewidth} }
\end{tabular}
\caption{Varying the number of query keywords} \label{fig:keywords}
\end{figure}

Next, we study the query performance when varying the number of
query keywords from 2 to 5. The results are presented in
Figure~\ref{fig:keywords}. The y-axes are also in logarithmic scale.
Again, our approach provides results with the best running time over
all the three datasets, and it runs 1-2 times faster than the best
baseline, IRT. For IF, we observe that the more keywords are
queried, the faster the results are returned. This is because IF has
more query keywords to do the filtering, and IF compute the match
distance for fewer trajectories that cover the query keywords.
For the other tree-based approaches, more query keywords require
more I/O cost to read the posting lists, and thus the running time
increases slightly.

%In addition, the time for computing the match distance for a
%candidate trajectory also increases with the increasing number of
%query keywords.

\subsubsection{Effect of partition granularity}
\begin{figure}[h!]
\centering \hspace{-14pt}
\psfig{figure=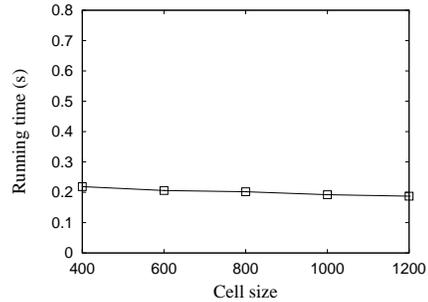,width=0.7\linewidth}
 \caption{Varying the number limit of trajectory segments in a
cell} \label{fig:quadcell}
\end{figure}

We now proceed to study the query performance of the proposed index
with regard to the partition granularity. Recall that we set a limit
for the number of trajectory segments in each cell. A cell splits
into $4$ sub-cells when the number of segments exceeds the limit. In
this experiment, we vary the number limit from $400$ to $1200$. The
results of running time and I/O cost are shown in
Figure~\ref{fig:quadcell}. From the figure, we can conclude that our
approach is not sensitive to the partition granularity. With finer
partition, the performance slightly degrades because more cells are
scanned but few additional trajectories are pruned. However, this
performance degradation is so small that it is negligible. In
particular, when varying the limit from $400$ to $1200$, the running
time degradation is only $0.03$s.

\subsubsection{Scalability}
\begin{figure}[h!]
\centering
\begin{tabular}{c}
\hspace{-5pt} \subfigure[Varying the number of trajectories]{
\psfig{figure=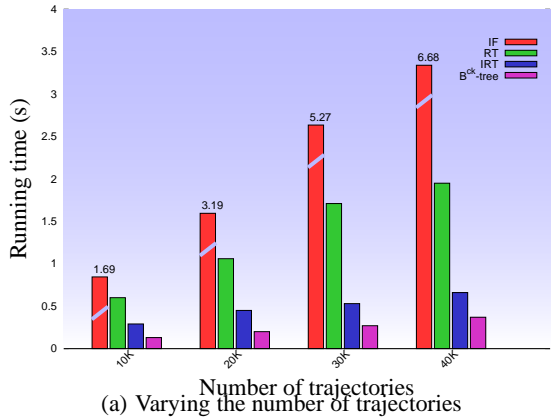,width=0.9\linewidth} \label{fig:scalability_traj}} \\
\hspace{-5pt} \subfigure[Varying the number of locations]{
\psfig{figure=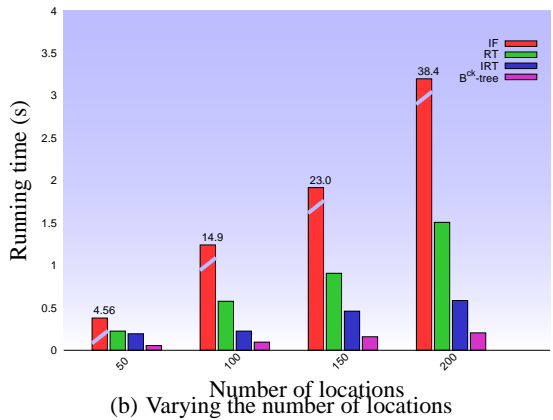,width=0.9\linewidth}
\label{fig:scalability_loc}}
\end{tabular}
\caption{Scalability results} \label{fig:scalability}
\end{figure}

Finally, we evaluate the scalability. In this experiment, we report
two sets of results. In the first set, we fix the number of
locations in each trajectory at 50 and vary the number of
trajectories from $10$K to $40$K. In the second one, we use one
datasets with $20$K trajectories and vary the number of locations in
each trajectory from $50$ to $200$. The running times are shown in
Figure~\ref{fig:scalability}. As expected, all of the four methods
take linear/sublinear time.  We also notice that the proposed method
B$^{ck}$-tree scales much better than do the other methods when
increasing the number of trajectories.

%The reason could be that the proposed method

%==========================================
\section{related work}\label{sec:related}
%==========================================

\stitle {Trajectory Query}\\
To the best of our knowledge, no work has considered answering the
proposed TkSK queries for trajectory data.

Related to the TkSK queries is the keyword range queries supported
in some online GPS trajectory sharing applications, e.g., Mountain
Bike (www.bikly.com), GPS sharing, etc., in which users can share,
browse and search GPS trajectories. They allow users to specify a
region and a set of keywords, and return the trajectories that are
inside the query region and contain the set of query keywords.
However, the algorithms used are not publicized.

Existing work on spital-temporal trajectory indexing
schemes~\cite{ssdbm99,PfoserJT00,tao01,
DBLP:conf/cidr/ChakkaEP03,DBLP:conf/icde/Cudre-MaurouxWM10} clearly
focuses on trajectories without text data. These index structures
are usually designed for keep tracking of moving objects. A number
of algorithms have been proposed to process different types of
spatial-temporal queries, such as $k$ nearest neighbor queries (
e.g., finding the k-closest objects with respect to a given point at
a given time), range queries ( e.g., finding all objects within a
given area), and complex spatial pattern queries~\cite{marios05,
marcos10, sigmod10}.

A number of similarity functions and algorithms have been developed
to compute the similarity between trajectories/time series data,
e.g.,~\cite{michail02,fodo,chen05}. Also, there exist work on
trajectory pattern discovery~\cite{miningtrajectory}, clustering
trajectories~\cite{clustertrajectory}, and finding significant
locations from trajectories~\cite{vldb10hot}.

\stitle {Spatial Keyword Search}\\
Zhou  et al.~\cite{weiying05} handle the problem of retrieving web
documents relevant to a keyword query within a pre-specified spatial
region. Similar problem is also considered by Chen  et
al.~\cite{chen06} and Hariharan  et al.~\cite{ssdbm07}.
These proposals use loose combinations of an inverted file and a
spatial index (e.g., R-tree).
 The
query processing in these proposals occurs in two stages: One type
of indexing (e.g. inverted list) is used to filter web document in
the first stage, and then the other index (e.g. R-tree) is employed,
or the vice versa.
This index has the disadvantage that it cannot simultaneously prune
the search space using both keywords and spatial distance.

Felipe et al.~\cite{Felipe08} propose a novel index structure called
\textit{IR$^2$-tree} that augments an R-tree with signatures.
%Each
%leaf entry $p$ stores a signature $p.\alpha$ as a fixed-length
%bitmap that summarizes the set of keywords of $p$. Each non-leaf
%entry $e$ stores a signature $e.\alpha$ that is the bitwise-OR of
%the signatures of the entries in the child nodes of $e$.
%
For the first time, the new hybrid index structure enables to
utilize both spatial information and text information prune search
space at query time, which advances the state-of-the art in
spatial-keyword query processing.
However, this proposal suffers from the crucial limits of signature
files (e.g., the number of false matches is linear in the collection
size~\cite{Zobel}).
Further, the \textit{IR$^2$-tree} faces the challenge of whether the
signatures possess enough pruning power to offset the extra cost
incurred by the taller trees that result from inclusion of
signatures.

The hybrid index structure that combines R$^*$-tree and bitmap
indexing is developed to process a new query called $m$-closest
keyword query~\cite{icde09} that returns the closest objects
containing at least $m$ keywords. This index structure exhibits the
same problems as do signature-file based indexing~\cite{Felipe08}.

The hybrid index structure IR-tree~\cite{vldb09} that integrates the
R-tree and inverted file enables the efficient processing of the
location-aware top-$k$ ranking query by utilizing both location and
text information to prune the search space.
In the IR-tree \cite{vldb09} %differs from the other augmented trees
%in two ways. First,
the fanout of the tree is independent of the number of words of
objects in the dataset, and, during query processing, only (a few)
posting lists relevant to the query keywords need to be fetched.
%
%
%A new index structure called Spatial-Keyword Inverted File is
%proposed to handle location-based web
%searches~\cite{DBLP:conf/dexa/KhodaeiSL10}, which transform
%
A recent proposed index named Spatial Inverted
Index~\cite{DBLP:conf/ssd/RochaGJN11}  maps each keyword  to a
distinct aggregated R-tree~\cite{Papadias:2001} that stores the
objects containing the given keyword.
The collective spatial keyword
query~\cite{DBLP:conf/sigmod/CaoCJO11} aims to retrieve a group of
nearby objects that cover the query keywords.

None of these proposals considers trajectory data associated with
text as does this paper.
Moreover, these proposed hybrid index solutions are not supported by
the mainstream DBMSs. In contrast, the proposed solution in this
paper is ready to be implemented on the DBMSs.

% to be added

Finally, note that the proposed T$k$SK  query is complementary to
the route planning queries(e.g.,~\cite{Sharifzadeh:2008}), which
return a route of places from a spatial database such that the route
covers a set of query keywords and the travel distance is minimized.

%==========================================
%\section{Conclusions and Future Work}\label{sec:conc}
\section{Conclusion}\label{sec:conc}
%==========================================
This paper proposes a new algorithm IE for efficiently answering
T$k$SK queries on trajectory data associated with text descriptions.
The algorithm is developed based on a new hybrid index called
cell-keyword conscious B$^+$-tree, denoted by \cellbtree. \cellbtree
allows us to develop algorithms that exploit both text relevance and
location proximity to facilitate efficient and effective query
processing. Additionally, the algorithm Match is proposed for
efficiently computing the match distance between a query and a
trajectory. The experimental results demonstrate that the proposed
algorithm outperforms several baseline algorithms significantly and
offers good scalability.

%meihui: remove the follows and expand the conclusion a bit more on contribution.

%Several promising directions for future work exist. First, it will
%be interesting to process other types of queries for trajectory data
%where keywords using the proposed index. For example, the proposed
%algorithm IE can be  employed to find trajectories that contain a
%set of keywords and fall in a given region, and we will study the
%performance. Second, the proposed index \cellbtree can be extended
%for querying spatial Web objects (supported by many services, e.g.,
%Google Maps and Twitter) and it would be interesting to evaluate its
%performance.

\bibliographystyle{abbrv}
 \begin{small}
 \bibliography{ref}
 \end{small}

\normalsize

\appendix

\small
\section{Discussion on Updates}
 Deleting and replacing trajectories would seldom happen to a
trajectory repository. Thus we only consider inserting new
trajectories. To insert a new trajectory, the insertion algorithm
first locates those cells into which the trajectory falls. After
that, it checks whether the number of trajectory segments in each of
the located cells is still within the limit. If it is the case, the
algorithm associates the words of the trajectory with the
corresponding cells according to the method discussed in
Section~\ref{sec:alg:index} and inserts the entry $<$wordID, cellID,
tID$>$ into the $B^+$-tree. For a cell where the number of
trajectory segments exceeds the limit after insertion, the cell will
be split into 4 sub-cells. A re-computation of the word-cell
association is needed. After that, the algorithm inserts into the
$B^+$-tree the entries with respect to the newly created four
(sub-)cells and removes the obsolete entries associated to the old
cell.

%{\bf note}.. the following will be removed. but the two prunning may
%make the ITB-tree based solution a bit better.

\section{Baseline 4: ITB-tree Index Based algorithm}
\label{ssec:itb-tree}

%We briefly present the ITB-tree %in Section~\ref{app:itb-tree}, and an
% and an algorithm for processing T$k$SK query \PartialEval.

 % in
%Section~\ref{app:TBT}.

%\subsubsection{ITB-tree Index}\label{app:itb-tree}

The baselines TR and IRT treat each trajectory as an object to build
index. The ITB-tree index treats each location of a trajectory,
rather than the whole trajectory, as an object.

We proceed to briefly present an index structure, the ITB-tree
(Inverted file augmented TB-tree), and the idea of an algorithm
based on the ITB-tree for the T$k$SK query.
The ITB-tree is essentially a TB-tree~\cite{PfoserJT00} augmented
with inverted files.
The TB-tree~\cite{PfoserJT00} is proposed for indexing trajectory
data without text information to efficiently support location based
queries.
The ITB-tree inherits the property of TB-tree~\cite{PfoserJT00} that
 is capable of preserving consecutive locations of the same
trajectory in an index.
%
%A leaf node in an ITB-tree does not contain locations from different
%trajectories. Also, leaf nodes in an ITB-tree are connected via a
%doubly linked list if they contain consecutive parts of the same
%trajectory.
%

Each leaf node in the ITB-tree contains entries of the form
$e=(\Lambda, \psi)$, where $e$ represents a place of a trajectory in
dataset $\mathcal{D}$, $e.\Lambda$ is the minimum bounding rectangle
(MBR), which is a point for a place, and $e.\psi$ refers to the id
of the text description of the place. Each leaf node contains a
pointer to an inverted file with the text descriptions of the
objects stored in the node.
In addition, each leaf node maintains two pointers (forward and
backward) that link the leaf node to other leaf nodes that contain
adjacent sub-trajectories of the sub-trajectory contained in the
leaf node.

Each non-leaf node $CN$ in the ITB-tree contains a number of entries
of the form $(e, \lambda, \psi)$ where $e$ is the address of a child
node of $R$, $\lambda$ is the MBR of all rectangles in entries of
the child node, and $\psi$ is the identifier of a pseudo text
description that is the union of all text descriptions in the
entries of the child node. The pseudo text description is a union of
the text descriptions of the children nodes.
Each non-leaf node also contains a pointer to an inverted file with
the text descriptions of the entries stored in the node.

%
%\subsubsection{ Partial Query Evaluation Algorithm (\PartialEval)}
%\label{app:TBT}

%Based on \MatchDist, we present algorithm \PartialEval.

%We use the first layer of index to choose candidate trajectories
%that are likely to be a top-k result, and employ the second layer of
%index to compute the Match Distance for a candidate trajectory.

We treat query $q$ as a set of partial queries, where each partial
query has a keyword in $q.\psi$ and the spatial component
$q.\lambda$. For each partial query we find its nearest places
incrementally using the ITB-tree index.
When a trajectory is covered by all the partial queries, i.e., some
place in the trajectory is retrieved as a nearby place for each
partial query, we choose the trajectory as a candidate and compute
the match distance of the trajectory to the query.
Intuitively, the trajectory would be a good candidate of the top-$k$
results since it contains all the query keywords and its places
covering the keywords are close to the spatial component of the
query. The detailed pseudo code of the partial query evaluation
Algorithm
 can be found in our technical report.

\end{document}